\documentclass[a4paper,floatfix,superscriptaddress,pra,twocolumn,longbibliography]{revtex4-2}
\usepackage[final]{graphicx}
\usepackage{times,bbm,amsmath,amssymb}
\usepackage{epsfig,color}
\usepackage{xcolor}
\usepackage{hyperref}
\usepackage{adjustbox}
\hypersetup{
    colorlinks = true
}
\usepackage{cleveref}
\usepackage[utf8]{inputenc}
\usepackage{float,siunitx}
\usepackage[caption = false]{subfig}

\usepackage[greek,english]{babel}
\usepackage{thumbpdf,enumerate}
\usepackage{booktabs}
\usepackage{sidecap}
\usepackage[scaled=.8]{couriers}    
\usepackage{pstricks}
\usepackage{multirow}
\usepackage{placeins}
\usepackage{relsize}  
\usepackage{pst-grad,bm}
\usepackage{epigraph}
\usepackage{longtable}
\usepackage{soul}
\usepackage{dsfont}
\usepackage{ulem} 
\usepackage{acronym}
\usepackage{physics}

\begin{document}
\title{Polarization-encoded photonic quantum-to-quantum Bernoulli factory based on a quantum dot source}

\author{Giovanni Rodari}
\affiliation{Dipartimento di Fisica - Sapienza Universit\`{a} di Roma, P.le Aldo Moro 5, I-00185 Roma, Italy}

\author{Francesco Hoch}
\affiliation{Dipartimento di Fisica - Sapienza Universit\`{a} di Roma, P.le Aldo Moro 5, I-00185 Roma, Italy}

\author{Alessia Suprano}
\affiliation{Dipartimento di Fisica - Sapienza Universit\`{a} di Roma, P.le Aldo Moro 5, I-00185 Roma, Italy}

\author{Taira Giordani}
\affiliation{Dipartimento di Fisica - Sapienza Universit\`{a} di Roma, P.le Aldo Moro 5, I-00185 Roma, Italy}

\author{Elena Negro}
\affiliation{Dipartimento di Fisica - Sapienza Universit\`{a} di Roma, P.le Aldo Moro 5, I-00185 Roma, Italy}

\author{Gonzalo Carvacho}
\email{gonzalo.carvacho@uniroma1.it}
\affiliation{Dipartimento di Fisica - Sapienza Universit\`{a} di Roma, P.le Aldo Moro 5, I-00185 Roma, Italy}

\author{Nicol\`o Spagnolo}
\affiliation{Dipartimento di Fisica - Sapienza Universit\`{a} di Roma, P.le Aldo Moro 5, I-00185 Roma, Italy}

\author{Ernesto F. Galv\~{a}o}
\affiliation{International Iberian Nanotechnology Laboratory (INL), Av. Mestre José Veiga, 4715-330 Braga, Portugal}
\affiliation{Instituto de F\'{i}sica, Universidade Federal Fluminense, Niter\'{o}i -- RJ, Brazil}

\author{Fabio Sciarrino}
\affiliation{Dipartimento di Fisica - Sapienza Universit\`{a} di Roma, P.le Aldo Moro 5, I-00185 Roma, Italy}

\begin{abstract}

A Bernoulli factory is a randomness manipulation routine that takes as input a Bernoulli random variable, outputting another Bernoulli variable whose bias is a function of the input bias. Recently proposed quantum-to-quantum Bernoulli factory schemes encode both input and output variables in qubit amplitudes. This primitive could be used as a sub-routine for more complex quantum algorithms involving Bayesian inference and Monte Carlo methods. Here, we report an experimental implementation of a polarization-encoded photonic quantum-to-quantum Bernoulli factory. We present and test three interferometric set-ups implementing the basic operations of an algebraic field (inversion, multiplication, and addition) which, chained together, allow for the implementation of a generic quantum-to-quantum Bernoulli factory. These in-bulk schemes are validated using a quantum dot-based single-photon source featuring high brightness and indistinguishability, paired with a time-to-spatial demultiplexing setup to prepare input resources of up to three single-photon states.
\end{abstract}

\maketitle

\section{Introduction}

The so-called Bernoulli Factory (BF) problem was formally introduced by Keane and O'Brien in \cite{keane1994bernoulli}, extending a riddle posed by Von Neumann in \cite{neumann1951various} regarding the possibility of obtaining an unbiased coin starting from a biased one. In its classical formulation, the goal of a BF is to construct an algorithm which, starting from i.i.d. samples drawn from a Bernoulli variable with \textit{unknown} bias $p$, produces as output an exact realization of a Bernoulli variable with bias $f(p)$, for a given well-defined function $f: [0,1] \rightarrow [0,1]$. Since its original formulation, this BF problem has been extended \cite{leme2022multiparameter} and has found application in the development of several fields, such as Montecarlo methods \cite{flegal2012exact, vats2022efficient}, Bayesian inference \cite{gonccalves2023exact, herbei2014estimating} and mechanism design \cite{dughmi2021bernoulli, cai2021efficient}.

Given the broad applicability of such a problem, in recent years quantum counterparts to the BF problem were theoretically introduced \cite{Dale2015, Dale2016, Jiang2018} and experimentally implemented \cite{Patel2019, Zhan2020,liu2021general,Yuan2016}. In the quantum-to-classical version of a BF \cite{Dale2015,Dale2016}, the input classical coin is replaced by a \textit{quantum coin} represented by a two-dimensional state $\ket{p}$. In this mapping, the state $\ket{p}$ is taken in such a way that if it is measured in the computational basis, it returns a Bernoulli random variable with bias $p$. Interestingly, it was proven theoretically that with a quantum-to-classical BF one can implement a strictly larger set of output functions $f(p)$ than its classical counterpart, leading to a quantum advantage in terms of resources needed to perform such a task \cite{Dale2015,Patel2019}.
Stemming from this, the so-called Quantum-to-Quantum Bernoulli factory (QQBF) concept was introduced \cite{Jiang2018}. Here, both the inputs and the outputs of the protocol are quantum resources and, while it was shown that the set of quantum-to-quantum and classical-to-classical simulable functions have no inclusion relationship with each other \cite{Jiang2018}, the formal evaluation of any kind of advantage in terms of resources is, to our knowledge, still an open problem. Indeed, a direct comparison is made difficult exactly due to the lack of a proper inclusion relation between the classes of functions implementable by quantum and classical Bernoulli factories. Nonetheless, its fully quantum nature makes a QQBF suitable for use as an intermediate step within a broader quantum algorithm. With this in mind, a \textit{genuine} QQBF implementation should satisfy some requirements. First of all, the QQBF should be oblivious to the input bias $p$. That is, the scheme employed should be able to map - upon suitable postselection - an arbitrary, unknown state $\ket{p}$ into the state $\ket{f(p)}$. Secondly, the implementation should feature \textit{modularity}, i.e. the ability of chaining together several QQBF or integrating them as a processing step in a more general algorithm.
Most previous experimental demonstrations of QQBF \cite{Zhan2020,liu2021general} were essentially unable to fulfill these two requirements. 

In this work, we design and implement a genuine QQBF exploiting polarization-encoded photonic qubits and manipulating them within a full in-bulk interferometric setup. With this objective, we exploit a near-deterministic, high-brightness single-photon source (SPS) based on quantum dot (QD) technology, able to generate highly indistinguishable photons. Pairing it with a time-to-spatial demultiplexing setup, we can prepare the required 3-photon states to be injected into a modular implementation of a photonic QQBF with up to two stages, each performing one of the basic operations over a complex field. 
The paper is organized as follows: we start by describing the main elements required for the construction of a genuine quantum-to-quantum Bernoulli factory, and our proposed implementation using a photonic polarization-based encoding in a bulk optics setup. We validate the correct operation of the proposed modules by experimentally characterizing the output quantum state fidelity of the two primitive building blocks necessary for the construction of a general QQBF, implementing respectively the (anti-) product and the arithmetic (harmonic) mean functions. Finally, we perform a three-photon experiment in which the proposed interferometers are concatenated together in a modular fashion. This allows us to construct a QQBF performing a more complex two-stage function, which can be seen as an instance of a programmable quantum device.

\subsection{Photonic quantum-to-quantum Bernoulli factory}

\begin{figure}[ht]
        \centering
        \includegraphics[width=1\columnwidth]{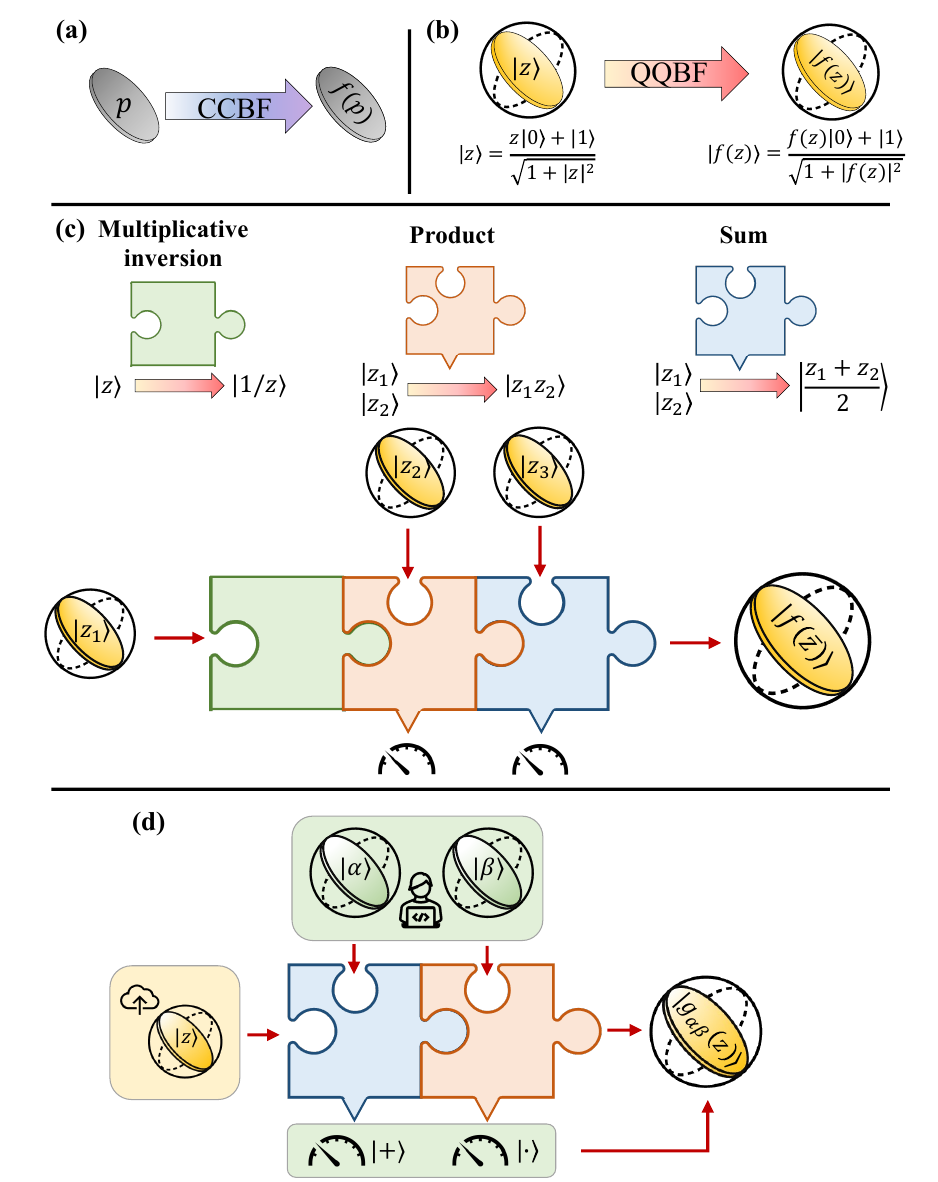}
        \caption{\textbf{Main properties of a Bernoulli factory construction.} (a) Classical-to-Classical Bernoulli factory (CCBF). A classical coin with bias $p$ is converted into an output classical coin with bias $f(p)$. (b) Quantum-to-Quantum Bernoulli factory (QQBF). A quantum coin $\ket{z}$ is used as input and manipulated into a quantum state $\ket{f(z)}$. (c) Modularity of a QQBF. A QQBF implementing a generic rational function $f(z)$ on the complex field can be obtained by concatenating a set of the three primitive building blocks in a modular fashion: the multiplicative inversion, the product, and the sum. (d) Quantum programmability of a QQBF. Given an unknown input quantum coin state, a user can prepare particular quantum states to program the functional shape $g_{\alpha \beta}(z)$ of the transformation to be performed by the QQBF.}
        \label{fig:sch}
\end{figure}

As briefly presented in the introduction, the quantum-to-quantum Bernoulli factory problem is an extension of its classical counterpart \cite{keane1994bernoulli}, in which both input and output coins are encoded in quantum states \cite{Jiang2018} (Fig. \ref{fig:sch}a-b). More specifically, given a generic complex number $z \in \mathbb{C}$ we can define the input state as 

\begin{equation}
    \ket{z} = \frac{z\ket{0_L}+\ket{1_L}}{\sqrt{1+\abs{z}^2}}
\end{equation}
which is the stereographic projection of the Bloch sphere onto the complex plane. With this mapping, apart from an irrelevant phase factor, a state $\ket{z}$ when measured in the computational basis returns a Bernoulli distribution with bias $p \in \mathbb{R}$ given by:
\begin{equation}
    p = \frac{1}{1 + \abs{z}^2}
\end{equation}
The goal of a QQBF is to process a set of quantum coins in the same unknown initial state $\ket{z}$, to return as the final state an output quantum coin in the form:
\begin{equation}
    \ket{f(z)} = \frac{f(z)\ket{0_L}+\ket{1_L}}{\sqrt{1+\abs{f(z)}^2}},
\end{equation}
where $f(z)$ is a chosen function. 
In Ref. \cite{Jiang2018} the necessary and sufficient condition for the existence of a quantum-to-quantum Bernoulli factory $\ket{z} \rightarrow \ket{f(z)}$ were found, namely that the function $f$ belongs to the set of rational functions in the parameter $z$ on the complex field.  
Note that it can be proven that each function $f(z)$ belonging to the set of rational functions over the complex field can be obtained by simply concatenating the operations of a field, i.e. multiplicative inverse, product and sum, as depicted in (Fig. \ref{fig:sch}c).
This means that an arbitrary quantum-realizable QQBF can be constructed providing that each primitive field operation can be implemented on its own, and concatenated in a \textit{modular} fashion.

In a photonic formalism fully relying on linear-optical elements \cite{kok2007linear,slussarenko2019photonic,flamini2018photonic}, practically speaking, the modularity requirement can be enforced in a \textit{bias-oblivious} way, i.e. with no information on the input state parameters, and by encoding both input and output states in a specific photonic degree of freedom, which can be manipulated without affecting the others. 
We note that in such a modular implementation, a QQBF can also be understood as an instance of a programmable quantum device \cite{nielsen1997programmable, duvsek2002quantum, rovsko2003generalized, ziman2005realization}, a paradigm in which a transformation or a measurement performed upon a set of \textit{data} quantum states is controlled via input \textit{program} qubits provided by the user. Programmable quantum devices are at the core of delegated blind quantum computing schemes \cite{fitzsimons2017unconditionally, Polacchi2023}. Under this perspective, conceptually depicted in Fig. \ref{fig:sch}d, the transformation implemented by the QQBF can be chosen and controlled by an external user who provides to the QQBF a set of chosen program quantum registers $\{\ket{\alpha_1}, \ket{\alpha_2}, \ldots \}$. As an example, in this work we implement a QQBF able to manipulate a single unknown input quantum coin $\ket{z}$ into an output quantum coin $\ket{g_{\alpha_1\alpha_2\ldots}(z)}$, in such a way that the functional form of the bias $g(\cdot)$ is determined by the parameters in the program states. That is, in a 3-photon scenario given two program states $\ket{\alpha}, \ket{\beta}$ one can implement a QQBF that transforms the bias of unknown input qubit $\ket{z}$ according to any chosen linear function:
\begin{equation}
    \ket{z} \otimes \ket{\alpha}\ket{\beta} \; \rightarrow \; \ket{\alpha z + \beta} \equiv \ket{g_{\alpha \beta}(z)}
    \label{eqn:program}
\end{equation}
A more general description of the space of functions achievable via a \textit{genuine} implementation of a QQBF protocol is reported in the Supplementary Materials.

In literature, different experiments have attempted to realize a QQBF using photonic-based platforms, but they fail in proposing a general approach in which different operations can be concatenated without any knowledge about the input quantum state \cite{liu2021general,Zhan2020}. 

We note that recent advances in photonic technologies \cite{wang2020integrated} have enabled the realization of system setups comprising several optical spatial modes and components, enabled by the properties and the encoding schemes of integrated platforms \cite{flamini2015thermally,clements2016optimal,taballione2021universal,maring2023general}. However, spatial-mode based encoding schemes may be significantly challenging to implement and stabilize for long-distance communication. Therefore, towards a practical integration of a quantum-to-quantum Bernoulli factory as a primitive within quantum networks, it is necessary to employ a different noise-resilient encoding for long distance transmission, such as the polarization degree of freedom. While conversion of spatial-mode encoded states to polarization states has been previously achieved on silicon based on-chip platforms \cite{ma2016silicon, llewellyn2020chip}, in this context another approach involves devising protocols that directly employ polarization. With this in view, here we propose an implementation of a genuine quantum-to-quantum Bernoulli factory protocol based on in-bulk optical elements, which employs the polarization of distinct photons to encode the input quantum coins $\ket{z}$. In particular, we propose and experimentally characterize the performance of a pair of versatile interferometric setups, which exploit bosonic interference effects among photons to implement the set of primitive field functions $\ket{f(z)}$ required for the construction of a general polarization-based QQBF.

\section{Results}
\subsection{Polarization-based implementation of a quantum-to-quantum Bernoulli Factory}

\begin{figure*}[ht]
    \centering
    \includegraphics[width=1\textwidth]{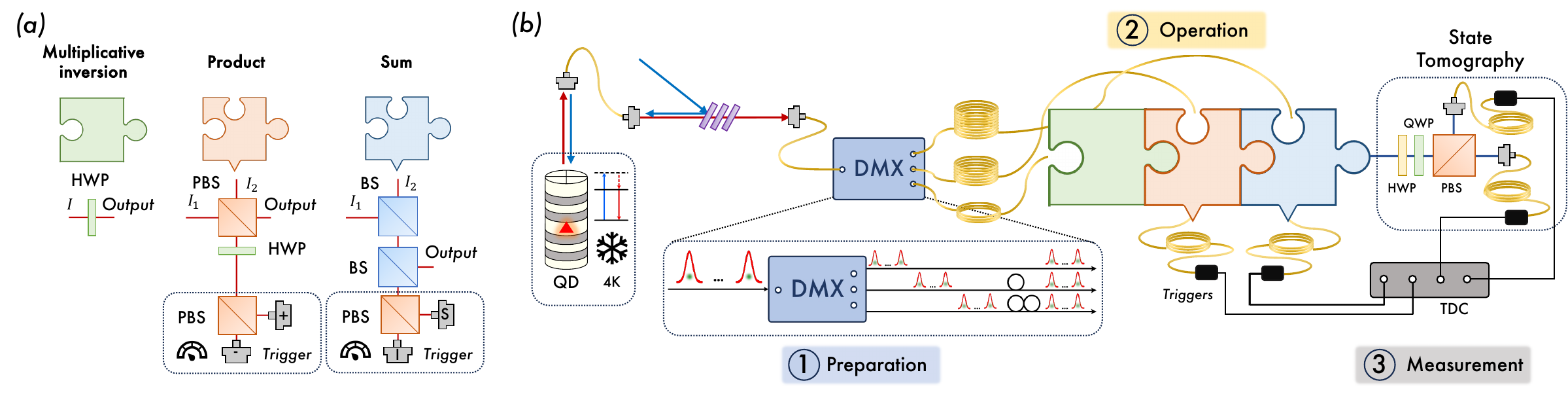}
    \caption{\textbf{Polarization-encoded Quantum-to-Quantum Bernoulli factory (QQBF).} \textit{(a) Primitive building blocks of a polarization-based QQBF:} three full in-bulk interferometers implementing the primitive field operations required for a generic QQBF. The input modes of the interferometers are labelled $I_1$ or $I_2$. These modules can be concatenated to construct more general functions in a \textit{modular} approach visualized here as puzzle pieces. \textit{(b) Full experimental apparatus:} We employ a quantum dot single photon source kept in a cryostation at 4K and operated non-resonantly in the so-called longitudinal acoustic (LA) phonon-assisted configuration \cite{thomas2021bright}. 
    The output of the QD source is connected to a time-to-spatial demultiplexing setup (DMX) which distributes the single photon stream in bunches of $\sim 180$ ns towards three output modes: here, temporal synchronization is achieved via finely tuned in-fiber delay loops. The three-photon resource states are then input to the modular QQBF setup: at its output, the set of projective measurements required for state reconstruction is performed with a sequence of quarter-, half-waveplates, and PBS. Photon counting events recorded with avalanche photodiodes are processed in a time-to-digital converter where two- or three-fold coincidence events with the trigger outputs, denoting that the protocol have succeeded, are suitably postselected. \textit{Legend:} PBS: Polarizing beam splitter, BS: Balanced beam splitter, HWP: Half-wave plate, QWP: Quarter-wave plate, $+/-$ and $\text{S}/\text{I}$: single-photon detectors used, respectively, for product/antiproduct and arithmetic mean/harmonic mean.
    }
    \label{fig:exp_apparatus}
\end{figure*}

In this section we demonstrate how the basic operations of a QQBF - inversion, product, sum - can be implemented using polarization of single photons qubits in an in-bulk, photonic linear-optical inteferometric set-up. 
As commonly done in a polarization-based encoding, we identify the horizontal/vertical basis $\{\ket{H}, \ket{V} \}$ of the photon polarization as the computational basis $\{\ket{0_L}, \ket{1_L}\}$. Consequently, the generic state of a quantum coin $\ket{z}$ can be written as:
\begin{equation}
    \ket{z} = \frac{z\ket{H}+\ket{V}}{\sqrt{1+\abs{z}^2}}
    \label{eqn:mapping}
\end{equation}
With this parameterization, the primitive field operations can be implemented using the interferometers depicted in Fig.~\ref{fig:exp_apparatus}a, which we now briefly describe.

\textbf{Multiplicative inversion operation}. The inversion operation, defined as $\ket{z} \rightarrow \ket{\frac{1}{z}}$, is performed via a half-wave plate (HWP) with the optical axis placed at an angle of $ \pi/4$ radians with respect to the horizontal axis. This is the only operation in the protocol which is deterministic and acts on only one photon.

\textbf{Product operation}. The implementation of the product operation, defined as $\ket{z_1}\ket{z_2} \rightarrow \ket{z_1z_2}$, requires two photons, with states $\ket{z_1}$ and $\ket{z_2}$, entering into the two input ports of a polarizing beam splitter (PBS). Post-selecting only the events with a single output photon per PBS output mode, we then measure the state of the trigger photon to select the target photon's correct state. Acting on the trigger photon with a HWP with an optical axis rotated by an angle of $\pi/8$ and a PBS, if its polarization is found to be $\ket{V}$ ($+$ output) the state of the target photon is the product between $z_1$ and $z_2$ ($\ket{z_1 z_2}$), while if the polarization is found to be $\ket{H}$ ($-$ output) the target output state is the antiproduct ($\ket{-z_1 z_2}$). In principle, it is possible to always produce the product output $\ket{z_1 z_2}$ by applying a fast polarization rotation, e.g. by means of an electro-optical modulator, acting conditioned upon a detection on the ``$-$" output to recover the correct phase relation.
In our implementation the experiment is performed under post-selection conditions, meaning that the probability of success of the product $P_+$ and the probability of the antiproduct $P_-$ are given respectively by:
\begin{equation}
    P_+ (z_1,z_2)= P_- (z_1,z_2) = \frac{1+\abs{z_1}^2\abs{z_2}^2}{2(1+\abs{z_1}^2)(1+\abs{z_2}^2)}
\end{equation}
whose maximum $P_{+/-}^{max}=0.5$ is attained for input quantum coins satisfying $\abs{z_1}^2 = \abs{z_2}^2 = 0$. Moreover, as discussed in the Supplementary materials, we have that the success probability becomes null when one has $\ket{z_1} \rightarrow \ket{0}$ and $\ket{z_2} \rightarrow \ket{\infty}$, corresponding to a point where the product operation takes an indefinite form.

\textbf{Sum operation}. To implement the sum operation, defined as $\ket{z_1}\ket{z_2} \Rightarrow \ket{\frac{z_1+z_2}{2}}$, a pair of photons in states $\ket{z_1}$ and $\ket{z_2}$ enter the two input ports of a 50:50 beam splitter (BS). Only the events in which the two photons exit from the same output port are selected. Subsequently, this pair of photons are eventually split in two optical modes through a second BS and the polarization of the photon exiting the trigger output is measured in the computational basis. If the polarization is found to be $\ket{V}$ ($S$ output), then the output state of the target photon is the arithmetic mean ($\ket{\frac{z_1+z_2}{2}}$). Conversely, if the trigger's polarization is found to be $\ket{H}$ ($I$ output) the output state is the harmonic mean ($\ket{\frac{2 z_1 z_2}{z_1+z_2}}$). The exact sum operation $\ket{f(z_1, z_2)} = \ket{z_1 + z_2}$, i.e. without the additional factor two, can be retrieved simply by concatenating with this setup a product block multiplying the output state by this factor.
Again, due to the post-selection process, the success probability of the arithmetic mean $P_S$ is given by:
\begin{equation}
    P_S (z_1,z_2)= \frac{\abs{\frac{z_1+z_2}{2}}^2+1}{4(1+\abs{z_1}^2)(1+\abs{z_2}^2)}
\end{equation} 
while the probability of the harmonic mean $P_I$ is expressed as:
\begin{equation}
    P_I (z_1,z_2)= \frac{\abs{z_1+z_2}^2+\abs{2z_1 z_2}^2}{16(1+\abs{z_1}^2)(1+\abs{z_2}^2)}
\end{equation}
whose maximum is analytically found to be $P_{S/I}^{max}=0.25$. Again, we have some critical points where $P_{S/I}$ becomes null: $(z_1, z_2) = (\infty, \infty)$ for the sum operation and $(z_1, z_2) = (0, 0)$ for the harmonic mean operation, as discussed in more detail in the Supplementary Materials.

\textbf{Concatenation}.
To devise an optical setup implementing a more complex function, i.e. a generic rational function defined over the complex field, it is essential to be able to concatenate the basic operations. The fact that both input and output photonic qubits are encoded in polarization facilitates this concatenation.

Indeed, the output from each single building block operation can be used as the input of the next operation, with a clear distinction between post-selected and output modes of each building block. 
While the success probability will decrease exponentially with the number of building blocks used, it can be proven that any complex operation can be implemented with a finite number of steps. More precisely, to implement a rational function of degree $n$, the number of required steps is upper bounded by $4n$ (see the Supplementary Materials for a formal demonstration). Furthermore, we remark that finite success probabilities are a feature of general Bernoulli processes, and not a limitation that is specific to our proposal. 

\subsection{Experimental verification of the polarization based QQBF}

In this section we provide a description of the complete experimental apparatus that we employed for the verification of the proposed polarization-based modular QQBF scheme. 
As we briefly summarise in Fig.~\ref{fig:exp_apparatus}b, the full experimental setup can be divided into three stages: state preparation, operation, and measurement. As described in the figure, we employ a quantum dot (QD) \cite{gazzano2013bright, nowak2014deterministic, somaschi2016near, ollivier2020reproducibility, thomas2021bright} as a highly efficient, near-deterministic source of single photons.
At the output of the QD source, which we operate in the so called non resonant longitudinal acoustic (LA) phonon-assisted scheme \cite{thomas2021bright} at a repetition rate of $79$ MHz, we measure a single photon emission rate of around $\sim 3.5$ MHz on avalanche photodiodes. We can assess the quality of the single photon emission via a standard Hanbury-Brown-Twiss setup, measuring a second order autocorrelation of around $g^{(2)}(0) \sim 2 \%$; while the indistinguishability between subsequently emitted photons evaluated in a Hong-Ou-Mandel interference experiment \cite{hong1987measurement} is measured to be $V_{\text{HOM}} \sim 92 \%$, as shown in the Supplementary Materials.
The output of the QD source is connected via a single mode fiber to a time-to-spatial demultiplexing setup (DMX), similar to the one described in \cite{pont2022high,pont2022quantifying}, to distribute the single photon stream towards three output modes. To achieve up the three-photon resource states employed in the present QQBF implementation, time synchronization amongst photons in the three modes is obtained via in-fiber delay loops of different lengths.

The emission properties of the QD source, characterized by two degenerate energy levels each emitting H/V polarized photons in a sligthly asymmetric microcavity \cite{ollivier2020reproducibility}, prevent us from achieving a fully polarized single photon signal at the output of the source. 
Due to this fact, to achieve full control on the input state preparation, we first select the horizontally polarized component by means of a PBS. Subsequently, we employ a pair of HWP and QWP to encode arbitrary states in the polarization of single photons, as in the mapping given by Eq.\eqref{eqn:mapping}. After preparing the polarization of the photon states as $\ket{z_i}$, we use them as input in the desired interferometer implementing a primitive QQBF function, as introduced in the previous section. Practically, we note that we had to introduce either an additional HWP or a liquid crystals retarder to compensate for optical phases introduced upon reflection by the mirrors and PBS/BS employed in the interferometric setup. As mentioned previously, we post-select only events in which one of the two input photons is collected in the trigger path and the other in the output path. 
Performing the appropriate measurements on the trigger photon, we ensure that the state of the other photon - the one in the output arm - has been mapped into the desired QQBF result. Upon this post-selection process, in the last stage of the apparatus, we are able to perform quantum state tomography of the output photons' state by means of polarization projections implemented via a QWP, a HWP, and a PBS. 
Finally, to demonstrate the approach's modularity, we concatenate the two setups: the output of the sum block becomes one of the inputs of the product block, while the third photonic resource is taken from the previously unused channel of the DMX. In this setup, exploiting the high intrinsic brightness of the single photon source paired with the time-to-spatial demultiplexing setup, data acquisition is carried out on three-photon coincidence events. Indeed, in this case, the set of three projective measurements ($\sigma_x$,$\sigma_y$,$\sigma_z$) required to perform quantum state tomography, and verify that the state of the output photon is the desired one by computing quantum state fidelity, must be post-selected upon detection of two photons on a given combination of two trigger outputs.
 
\begin{figure*}[ht]
    \centering
    \includegraphics[width=1\textwidth]{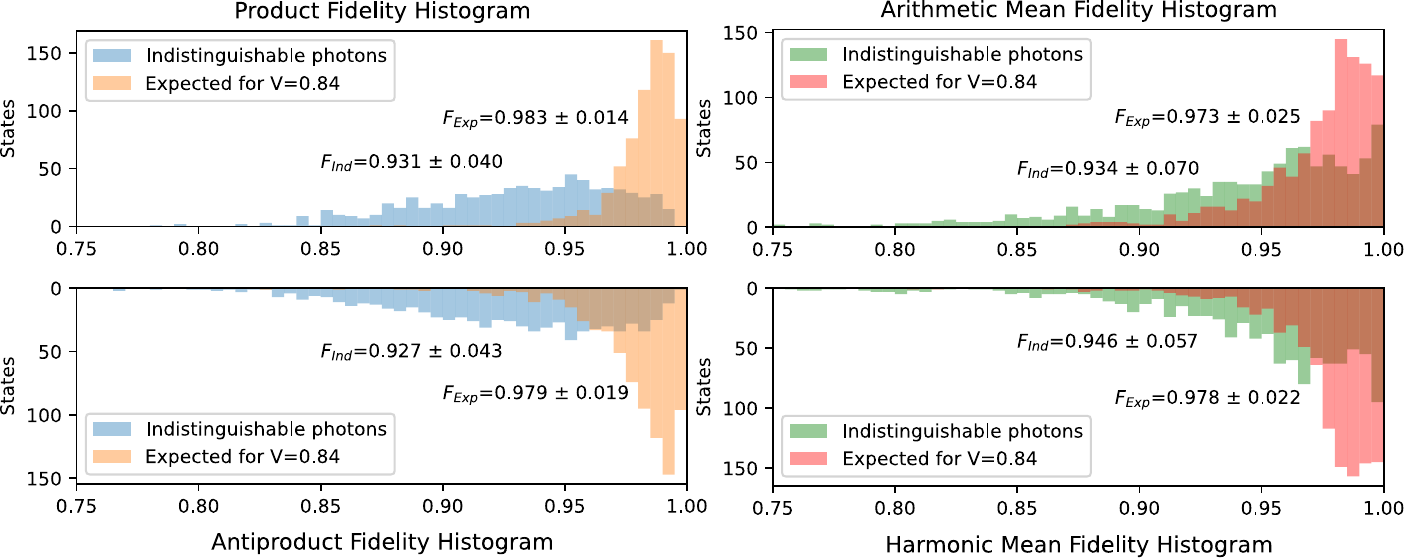}
    \caption{\textbf{Fidelity of the output states from the basic operation interferometers}. On the left side, histograms of the fidelity for product and antiproduct operation evaluated over $737$ random pairs of states $\ket{z_1}$ and $\ket{z_2}$ uniformly  drawn from the Bloch sphere. On the right, histograms of the fidelity for the arithmetic and harmonic mean operation evaluated over $1007$ random pairs of states $\ket{z_1}$ and $\ket{z_2}$ uniformly drawn from the Bloch sphere. The fidelities are shown both with respect to the ideal output pure state $\ket{f(z)}$, and with respect to the state expected when taking into account the estimated imperfect photonic indistinguishability. }
    \label{fig:prod_sum_common}
\end{figure*}

\subsubsection{Validation of the Product building block}

We first consider the primitive building block implementing the product operation as a quantum-to-quantum Bernoulli factory. In our implementation, exploiting the polarization degree of freedom manipulated via in-bulk optical elements, two main noise-inducing effects must be accounted for: (i) The optical phase introduced by reflection through mirrors and the first polarizing beam splitter, which can be corrected by introducing a controlled phase shift between the $\ket{H}$ and $\ket{V}$ polarizations before the heralding measurement; and (ii) The imperfect indistinguishability between the employed single photons, i.e. in degrees of freedom different than the polarization, which introduces a degree of impurity in the product/antiproduct density matrix. Indeed, it can be shown that given a degree of indistinguishability $V$, the final density matrix of the product/antiproduct operations will respectively be:
\begin{equation}
    \rho_{\text{prod}} = \frac{1}{1 + |z_1 z_2|^2}\mqty ( |z_1 z_2|^2 & Vz_1z_2 \\
    Vz^*_1z^*_2 & 1),
\end{equation}

\begin{equation}
    \rho_{\text{antiprod}} = \frac{1}{1 + |-z_1 z_2|^2}\mqty ( |-z_1 z_2|^2 & -Vz_1z_2 \\
    -Vz^*_1z^*_2 & 1),
\end{equation}
as discussed in the Supplementary Materials.
To validate the experimental apparatus implementing the product operation, we performed tomography measurements of the states resulting from the product and antiproduct operation between states $\ket{z_1}$ and $\ket{z_2}$ randomly sampled from a uniform distribution of states defined over the Bloch sphere (see Supplementary Materials). We then evaluated the state fidelity \cite{jozsa1994fidelity}:
\begin{equation}
    \mathcal{F} = \Tr{\sqrt{\sqrt{\rho_{\text{exp}}} \sigma_{\text{th}} \sqrt{\rho_{\text{exp}}}}}^2
\end{equation}
between the experimentally obtained output state $\rho_{\text{exp}}$ and the expected output state $\ket{\sigma_{\text{th}}}$.
In Fig. \ref{fig:prod_sum_common} we report the results over $737$ pairs of states, obtaining mean fidelities with the expected output states $\ket{z_1z_2}$ and $\ket{-z_1z_2}$ of:
\begin{equation}
    \begin{split}
        F_{\text{prod}} = 0.931 \pm 0.040 ,\\
        F_{\text{antiprod}} = 0.927 \pm 0.043 .
    \end{split}
\end{equation}
These values can be further processed to take into account the imperfect indistinguishability between the employed photons. In particular, performing an Hong-Ou-Mandel experiment between the pairs of DMX output photons, we measured a pairwise two-photon visibility of around $V \sim 0.84$. Comparing the obtained experimental data with the expected output density matrices obtained from an ideal operation with input photons having this amount of partial indistinguishability, we find an output average fidelity of:
\begin{equation}
    \begin{split}
        F^{\text{Exp}}_{\text{prod}} = 0.983 \pm 0.014, \\
        F^{\text{Exp}}_{\text{antiprod}} = 0.979 \pm 0.019 .
    \end{split}
\end{equation}

\subsubsection{Validation of the sum building block}

Analogously to what has been discussed for the product operation, also in the sum operation we have to account for the presence of a polarization-dependent optical phase introduced by reflection on the beam splitters. Again, this effect can be corrected by placing a HWP in the first preparation arm and a  variable phase retarder in the output arm. Similarly to the product operation, we performed a quantum state tomography reconstruction of the states resulting from the arithmetic and harmonic mean operations between state pairs $\ket{z_1}$ and $\ket{z_2}$ randomly sampled from a uniform distribution of states defined over the Bloch sphere (see Supplementary Materials). We report in Fig. \ref{fig:prod_sum_common} the results over $1007$ pairs of states, obtaining average fidelities with the expected states $\ket{\frac{z_1 + z_2}{2}}$ and $\ket{\frac{2z_1z_2}{z_1 + z_2}}$ of:
\begin{equation}
    \begin{split}
        F_{\text{sum}} = 0.934 \pm 0.070 \\
        F_{\text{amean}} = 0.946 \pm 0.057 
    \end{split}
\end{equation}
respectively. Again, this result can be further processed to take into account the imperfect indistinguishability $V$ in the expected output state, where the formal dependence of the output density matrices with $V$ is reported in Supplementary Materials. By considering a two-photon visibility of around $V \sim 0.84$, we obtain the following average fidelities:
\begin{equation}
    \begin{split}
        F^{\text{Exp}}_{\text{sum}} = 0.973 \pm 0.025, \\
        F^{\text{Exp}}_{\text{amean}} = 0.978 \pm 0.022 .
    \end{split}
\end{equation}
We note that when considering the fidelities of the experimentally obtained states with respect to the theoretical model assuming perfect indistinguishability, for around $3\%$ of the output states we obtain values lower than $0.75$: this is due to the fact that, with the employed scheme, the corresponding input pairs $\{ \vert z_1 \rangle, \vert z_2 \rangle\}$ appear to be more sensitive to both an imperfect compensation of the polarization-dependent phase shift introduced by the first BS and imperfect indistinguishability. Indeed, when accounting for the latter effect, the corrected visibility for such states lies in all cases in the interval $(0.75,0.9)$, i.e. the tail of the overall distribution reported in Fig. \ref{fig:prod_sum_common}. 
Considering both the sum-block and the product-block, we obtain a high average fidelity with respect to randomly chosen input pairs $\{z_1, z_2\}$. We obtain an average fidelity with respect to the expected pure output states of $\bar{F} = (0.933 \pm 0.024)$, certifying the correct behaviour of the overall implementation of the QQBF building blocks without accounting for sources of experimental noise of the present implementation, i.e. non-ideal optical elements and imperfect photonic indistinguishability. Accounting for the latter effect in terms of a correction applied to the expected output density matrices, we obtain an average fidelity of $\bar{F}_{\text{Exp}} = (0.978 \pm 0.020)$, which is a benchmark of how we could expect the proposed optical setups to perform with the use of ideal photonic resources.

\subsubsection{Polarization based QQBF as a modular programmable setup}

\begin{figure*}[ht]
    \centering
    \includegraphics[width=\textwidth]{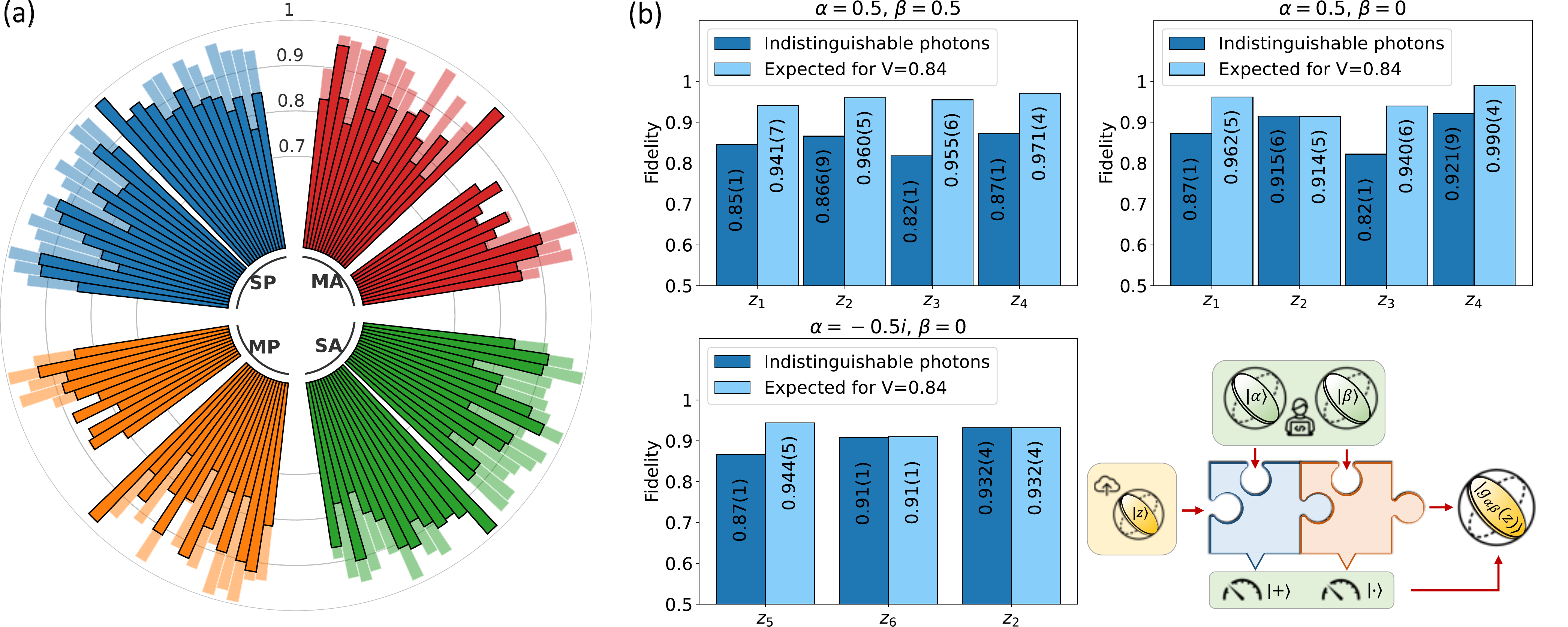}
    \caption{{\bf Performance of the concatenated QQBF primitive modules as a programmable device.} (a) Circular bar plot of the output quantum state fidelities, comparing the theoretical state with the one reconstructed via quantum state tomography, for each possible combination of concatenated operations. The black outlined bars represent the fidelities obtained with respect to the theoretically predicted pure output state; on top, the shaded bars represent the fidelity with respect to the state corrected for imperfect photonic indistinguishability. The absence of a bar corresponds to a null post-selection probability. (b) Interpretation of the results obtained by concatenating the sum and product primitive building blocks to implement a programmable QQBF producing the family of linear transformations $\ket{\pm g_{\alpha\beta}(z)} = \ket{\pm \alpha \pm \beta z}$. The three-bar plots depict the experimentally obtained fidelities when computing the output function $\ket{g_{\alpha\beta}(z)} = \ket{\alpha z + \beta}$ for different choices of the input data states $\ket{z}$, namely $\{z_1,z_2,z_3,z_4,z_5,z_6\}=\{1,0,-0.27-i0.74,-0.27+i0.74,-i,\infty\}$, and pairs of parameters $(\alpha, \beta)$. The program states at the input of the sum module and the product module, respectively, were chosen as $\ket{\beta/\alpha}$ and $\ket{2\alpha}$, so the pair of programming parameters are $(\alpha, \beta)=\{(1/2,1/2), (1/2,0), (-i/2,0)\}$ in the function $g_{\alpha \beta}(z)$. The darker and lighter blue bars show the quantum state fidelity between the experimentally reconstructed state and, respectively, the output state corresponding to completely indistinguishable input photons, and the state expected from an ideal operation with the experimentally characterized imperfect photonic indistinguishability.}
    \label{fig:concatenation}
\end{figure*}

As the natural next step, we then experimentally tested if the implemented in-bulk interferometers, discussed in the previous sections, could be chained together in a modular fashion via redirecting the output of a given block to the input of the subsequent one, an essential requirement towards the implementation of a photonic based QQBF implementing a generic function $\ket{f(z)}$. Indeed, by chaining together the sum and product modules depicted in Fig.~\ref{fig:exp_apparatus}a, one obtains a QQBF implementing, upon post-selection on the corresponding pair of trigger outputs, one of four possible functions $\ket{f(z_1,z_2,z_3)}$ among the three unknown input quantum coins $\{\ket{z_1},\ket{z_2},\ket{z_3}\}$: sum-product, harmonic mean-product, sum-antiproduct, and harmonic mean-antiproduct. From another perspective, we note that such a scenario, featuring two concatenated modules and three input photons, can be interpreted as the photonic realization of a programmable setup implementing the linear transformation of Eq. \eqref{eqn:program}. As a specific example, we identify the state $\ket{z_1}$ of one input photon as the data state $\ket{z}$, and treating the other two inputs $\{\ket{z_2}, \ket{z_3}\}$ as program states provided by the user. One can then realize both linear transformations given by:
\begin{equation}
    \ket{z} \otimes \ket{\alpha} \ket{\beta} \rightarrow \ket{\pm \alpha z \pm \beta}.
\end{equation}

In our implementation these functions are realized by choosing:
\begin{equation}
    \ket{z_1} = \ket{z}; \; \ket{z_2} = \ket{\frac{\beta}{\alpha}}; \; \ket{z_3} = \ket{2\alpha}
\end{equation}
since it can be easily shown that after the linear optical transformation obtained by chaining together the sum and product modules, the overall output state will be either in the state:
\begin{equation}
\ket{\frac{z + \beta / \alpha}{2} \cdot (2\alpha)} = \ket{\alpha z + \beta}
\end{equation}
or in the state:
\begin{equation}
\ket{- \frac{z + \beta / \alpha}{2} \cdot (2\alpha)} = \ket{-\alpha z - \beta}
\end{equation}
upon the post-selection process conditioned on the detection of one photon in the state $\ket{S}$ at the sum trigger output together with a photon at the product trigger output respectively in state $\ket{+}$ or in state $\ket{-}$.

To experimentally test the feasibility of this approach, we realized the interconnection between the two optical interferometers via a single-mode fiber, linking the sum output port to the input $I_1$ of the product interferometer in Fig.~\ref{fig:exp_apparatus}a. To compensate both for unwanted phases introduced by reflections in the product and sum setup, as well as the unchanged polarization state at the output of the sum setup, we introduce a set of paddles to compensate for the unknown unitary induced by the single-mode fiber on the polarization space.
Encoding different polarization states $\vert z_1 \rangle$, $\vert z_2 \rangle$ as input of the sum setup and $\vert z_3 \rangle$ as the free input in the product setup, we first evaluated the outcome of each of the four possible function combinations: sum-product, harmonic mean-product, sum-antiproduct and harmonic mean- antiproduct. The crucial use of a quantum dot based bright single-photon source paired with a highly efficient time-to-spatial demultiplexing setup enabled us to experimentally perform a three-fold quantum state tomography for each combination. In particular, we considered 30 different sets $\{ \vert z_1 \rangle, \vert z_2 \rangle, \vert z_3 \rangle \}$ of input states and for each of these we computed the fidelity associated with the four expected ideal output states, represented as black outlined bars in Fig \ref{fig:concatenation} (a). 
As briefly discussed in the previous sections, also in this scenario it is necessary to account for the noise introduced by the finite degree of indistinguishability. Indeed, as indicated by the shaded bars in Fig. \ref{fig:concatenation}a, the output state fidelity - our figure of merit for the experiment - generally improves when we consider the expected output states corrected for the limited degree of indistinguishability among the interfering single-photon states - assumed to be approximately $V \sim 0.84$, as reported in the Supplementary Materials. Note that there are triples of inputs $\{\vert z_1 \rangle, \vert z_2 \rangle, \vert z_3 \rangle \}$ for which the resulting density matrices do not depend on the visibility $V$, thus resulting in an output fidelity that is unaffected by this correction. Empty regions at the center of each bar plot indicate input choices for which the resulting function succeeds with null probability.
In Table \ref{Tab:fid3} we report the average fidelities, computed over the 30 input sets $\{\vert z_1 \rangle, \vert z_2 \rangle, \vert z_3 \rangle \}$, with respect to each possible combination obtainable by cascading together the sum and the product building blocks.
The average overall fidelity accounting for the visibility correction is found to be:
\begin{equation}
    \bar{F}^{\text{Exp}}_{\text{3-fold}} = 0.931 \pm 0.017 ,
\end{equation}
 which shows that the proposed polarization-based encoding and the fully in-bulk experimental schemes can be indeed chained together in a fashion that is both \textit{modular} and \textit{oblivious} about the input biases $\{\vert z_1 \rangle, \vert z_2 \rangle, \vert z_3 \rangle\}$, two essential requirements for the implementation of a \textit{genuine} version of a photonic quantum-to-quantum Bernoulli factory.

As briefly discussed at the beginning of this section, we can interpret the photonic concatenation of two QQBF primitive modules as an instance of a device implementing the family of linear functions $\ket{\pm g_{\alpha \beta}(z)} = \ket{\pm \alpha z \pm \beta}$. This perspective has one of the input photons interpreted as a data qubit $\ket{z}$, possibly unknown to the experimenter and provided by an external user; while the other two can be fixed by the experimenter to be $\ket{\beta / \alpha}$ and $\ket{2\alpha}$ so as to chose the parameters of the implemented function $g_{\alpha \beta}$ in a programmable fashion.
In Fig. \ref{fig:concatenation}b we focus on the results obtained for the QQBF implementing the function $\ket{g_{\alpha\beta}(z)} = \ket{\alpha z + \beta}$, recovered when post-selecting upon the detection of a photon in the $\ket{S}$-output of the sum module and on the $\ket{+}$-output of the product module. In particular, we report the experimentally obtained fidelities with respect to the (non) corrected output states for three different choices of programmable parameters $(\alpha, \beta)$, corresponding to $(\alpha = \beta = 1/2)$; $(\alpha = 1/2, \beta = 0)$ and $(\alpha = -i/2, \beta = 0)$. A comparison between the experimentally obtained density matrices and the expected ones is provided in the Supplementary Materials. Furthermore, by considering all the 30 chosen input triplets, we obtain an overall average corrected fidelity of $\hat{F}_{\alpha z + \beta} = 0.951\pm 0.024$ and $\hat{F}_{-\alpha z - \beta} = 0.945\pm 0.030$, showing that with our proposed polarization-encoded photonic setups we can provide a proof of principle realization of a programmable QQBF implementing the linear transformations $\ket{\pm g_{\alpha \beta}(z)} = \ket{\pm \alpha z \pm \beta}$.
\begin{table}[ht]
\begin{adjustbox}{width=\columnwidth,center}
\begin{tabular}{c|c|c|c|c}
$V$ &$F_{SP}$&$F_{MP}$&$F_{SA}$&$F_{MA}$\\
\hline
$1$& $0.865 \pm 0.055$&$0.851 \pm 0.061$&$0.866 \pm 0.059$&$0.849 \pm 0.058$\\
$0.84$ &$ 0.951 \pm 0.024$& $0.916 \pm 0.061$& $ 0.945 \pm 0.030$ & $0.913 \pm 0.059$
\end{tabular}
\end{adjustbox}
\caption{\textbf{Three-fold fidelity:} Average fidelities computed over 30 input triplets $\{ \vert z_1 \rangle, \vert z_2 \rangle, \vert z_3 \rangle \}$ for each of the four operation combinations (Sum-Product, Harmonic mean-Product, Sum-Antiproduct, Harmonic mean-Antiproduct). The reported values refer to the ideal case in which perfect indistinguishability is considered ($V=1$), and to the experimental   indistinguishability scenario ($V=0.84$). The first two table columns also represent the average fidelities for the photonic QQBF interpreted as a programmable setup, implementing the transformation $\ket{\pm g_{\alpha \beta}(z)} = \ket{\pm \alpha z \pm \beta}$.}\label{Tab:fid3}
\end{table}

\section{Discussion}

In this work we devised and experimentally verified a solution to the quantum counterpart of the so-called Bernoulli factory problem, i.e. a randomness manipulation protocol which generates, starting from quantum coins in a state $\ket{z} \propto z\ket{0} + \ket{1}$, an output state $\ket{f(z)}$ where $f(z)$ is a rational function defined on the complex field. In the present scheme the primitive field functions from which arbitrary QQBF can be constructed have been implemented by employing two fully in-bulk interferometric schemes, in which the required operations are realized as probabilistic gates acting on polarization-encoded photonic qubits. In particular, we experimentally tested the performance of the proposed interferometric setups both in the single-operation scenario and when two modules are concatenated, verifying not only the obliviousness of our schemes to the input state bias, but also the modularity of our approach, both requirements for a \textit{genuine} implementation of a QQBF. Moreover, we show how a QQBF where two or more modules are cascaded together can be interpreted as a programmable platform to perform an operation $g_{\alpha_1\alpha_2\ldots}(z)$ on an unknown input state $\ket{z}$, parameterized by a set of auxiliary program states $\ket{\alpha_1,\alpha_2,...}$, and presented an implementation of the function $\ket{\alpha z + \beta}$ as an example.
Our QQBF is implemented in a state-of-the-art hybrid photonic platform, comprised of a high-brightness single-photon source based on the quantum dot technology interfaced with an active time-to-spatial demultiplexing setup, tailored for the efficient generation of multi-photon resource states. This enables us to validate the present polarization-based photonic QQBF implementation via a reconstruction of the output state even in the more complex three-photon scenario. Moreover, even if the proposed scheme relies on bosonic interference effects and measurement post-selection thus being intrinsically probabilistic, one could take advantage of the near-deterministic properties of the QD emission to increase the final success probability of the protocol via the addition of a feedback-controlled active modulation system. 
This approach would provide an essential element towards the scalability of the proposed modular implementation of QQBF to perform complex functions requiring a higher number of modules chained together since the implementation of $n$ modules requires $n+1$ single-photon states. Given its probabilistic nature, a thorough analysis of the resource overhead of the protocol would be of interest to assess a practical application of our protocol, and is left as future work. Recent technological advances have shown the capability of moving photonic experiments towards regimes with larger photon numbers \cite{maring2023general, cao2024photonic, hansen2023single}. Hence, given the maturity of photonic quantum technologies, and considering that 
polarization encoded photonic qubits are a naturally suitable and resilient resource when transmitted over long distances, our results could pave the way toward the use of a QQBF, or similar randomness manipulation schemes, as a subroutine in a distributed computation over a complex quantum network.

\section{Materials and Methods}

Our quantum dot (QD) emitter is a single self-assembled InGaAs QD embedded at the center of an electrically controlled microcavity surrounded by two Braggs reflectors made of GaAs/Al$_{0.95}$Ga$_{0.05}$As $\lambda /4$ layers with 36 (16) pairs for the bottom (top) \cite{somaschi2016near}; to enhance the spontaneous emission via Purcell effect and enabling collection in a single mode fiber atop of the source. We employ a non-resonant emission geometry known as longitudinal acoustic (LA) phonon-assisted configuration \cite{thomas2021bright}: upon being excited with a 79 MHz-pulsed laser source at a wavelength of $927.2$ nm, photons are emitted non-resonantly at a wavelength of $927.8 \pm 0.2$ nm and separated from the residual pump laser through a sequence of three narrow bandpass filters. The QD source is housed on a commercially available solution - \emph{Quandela e-Delight} chip - and the single-photon source is kept at $4$K in a low-vibration closed-cycle He cryostat \emph{Attocube Attodry 800}. Pulse shaping of the 79MHz pulsed pump laser is obtained through a commercial 4f to select the required excitation central wavelength of $927.2$ nm and achieve a $\sim 100$ pm spectral bandwidth. As stated in the main text, the typical count-rate at the output of the SPS is around $3.5$ MHz when measured with avalanche photodiodes featuring a quantum efficiency of $\sim 35\%$, corresponding to a first lens brightness of $\sim 24\%$.
The stream of single photons emitted by the QD is then sent to the time-to-spatial demultiplexing setup. Here we have an optical setup that by means of the spatial diffraction induced by an acousto-optical modulator, divides the input signal in bunches of $\sim 180$ ns sequentially and periodically redirected towards three output channels, akin to the scheme employed in \cite{pont2022high, pont2022quantifying}. As stated in the main text, the signals from these outputs are collected in single-mode fibers, which are delay loops of different lengths in order to temporally synchronize the photon bunches exiting each of the three output channels, rendering them temporally indistinguishable.

\section{ACKNOWLEDGEMENTS}
This work was supported by the ERC Advanced Grant QU-BOSS (QUantum advantage via nonlinear
BOSon Sampling, grant agreement no. 884676) and by ICSC – Centro Nazionale di Ricerca in High Performance Computing, Big Data and Quantum Computing, funded by European Union – NextGenerationEU. E.F.G. acknowledges support from FCT - Fundação para a Ciência e a Tecnologia (Portugal) via project CEECINST/00062/2018.

\section{Author contributions}
F.H., T.G., G.C., N.S., E.F.G., F.S. conceived the scheme for the polarization-encoded Quantum-to-Quantum Bernoulli Factory. G.R., F.H., A.S. T.G., G.C., N.S., F.S. conceived the experimental platform. G.R., F.H., A.S. T.G., E.N., G.C., N.S., F.S. performed the experiment and carried out the data analysis. All authors contributed to discussing the results and to writing the manuscript.

\section{COMPETING INTERESTS}
F.H., T.G., G.C., N.S., E.G. and F.S. are listed as inventors on corresponding pending patent applications in Italy (No. 102023000012279) and Portugal (No. 20232005054430) both filed on 15th June 2023 and titled ‘Quantum Bernoulli Factory photonic circuit independent of input state bias’ dealing with schemes for the implementation of Quantum-to-Quantum Bernoulli factories. The authors declare no other competing interests.

\section{Data and materials availability}
All data needed to evaluate the conclusions in the paper are present in the paper and/or the Supplementary Materials.

%

\end{document}


\beginsupplement
\title{Supplemental Material: Polarization-encoded photonic quantum-to-quantum Bernoulli factory based on a quantum dot source}

\author{Giovanni Rodari}
\affiliation{Dipartimento di Fisica - Sapienza Universit\`{a} di Roma, P.le Aldo Moro 5, I-00185 Roma, Italy}

\author{Francesco Hoch}
\affiliation{Dipartimento di Fisica - Sapienza Universit\`{a} di Roma, P.le Aldo Moro 5, I-00185 Roma, Italy}

\author{Alessia Suprano}
\affiliation{Dipartimento di Fisica - Sapienza Universit\`{a} di Roma, P.le Aldo Moro 5, I-00185 Roma, Italy}

\author{Taira Giordani}
\affiliation{Dipartimento di Fisica - Sapienza Universit\`{a} di Roma, P.le Aldo Moro 5, I-00185 Roma, Italy}

\author{Elena Negro}
\affiliation{Dipartimento di Fisica - Sapienza Universit\`{a} di Roma, P.le Aldo Moro 5, I-00185 Roma, Italy}

\author{Gonzalo Carvacho}
\email{gonzalo.carvacho@uniroma1.it}
\affiliation{Dipartimento di Fisica - Sapienza Universit\`{a} di Roma, P.le Aldo Moro 5, I-00185 Roma, Italy}

\author{Nicol\`o Spagnolo}
\affiliation{Dipartimento di Fisica - Sapienza Universit\`{a} di Roma, P.le Aldo Moro 5, I-00185 Roma, Italy}

\author{Ernesto F. Galv\~{a}o}
\affiliation{International Iberian Nanotechnology Laboratory (INL), Av. Mestre José Veiga, 4715-330 Braga, Portugal}

\author{Fabio Sciarrino}
\affiliation{Dipartimento di Fisica - Sapienza Universit\`{a} di Roma, P.le Aldo Moro 5, I-00185 Roma, Italy}

\maketitle

\section{Source Characterization}
\begin{figure*}[ht]
    \centering
    \includegraphics[width=1\linewidth]{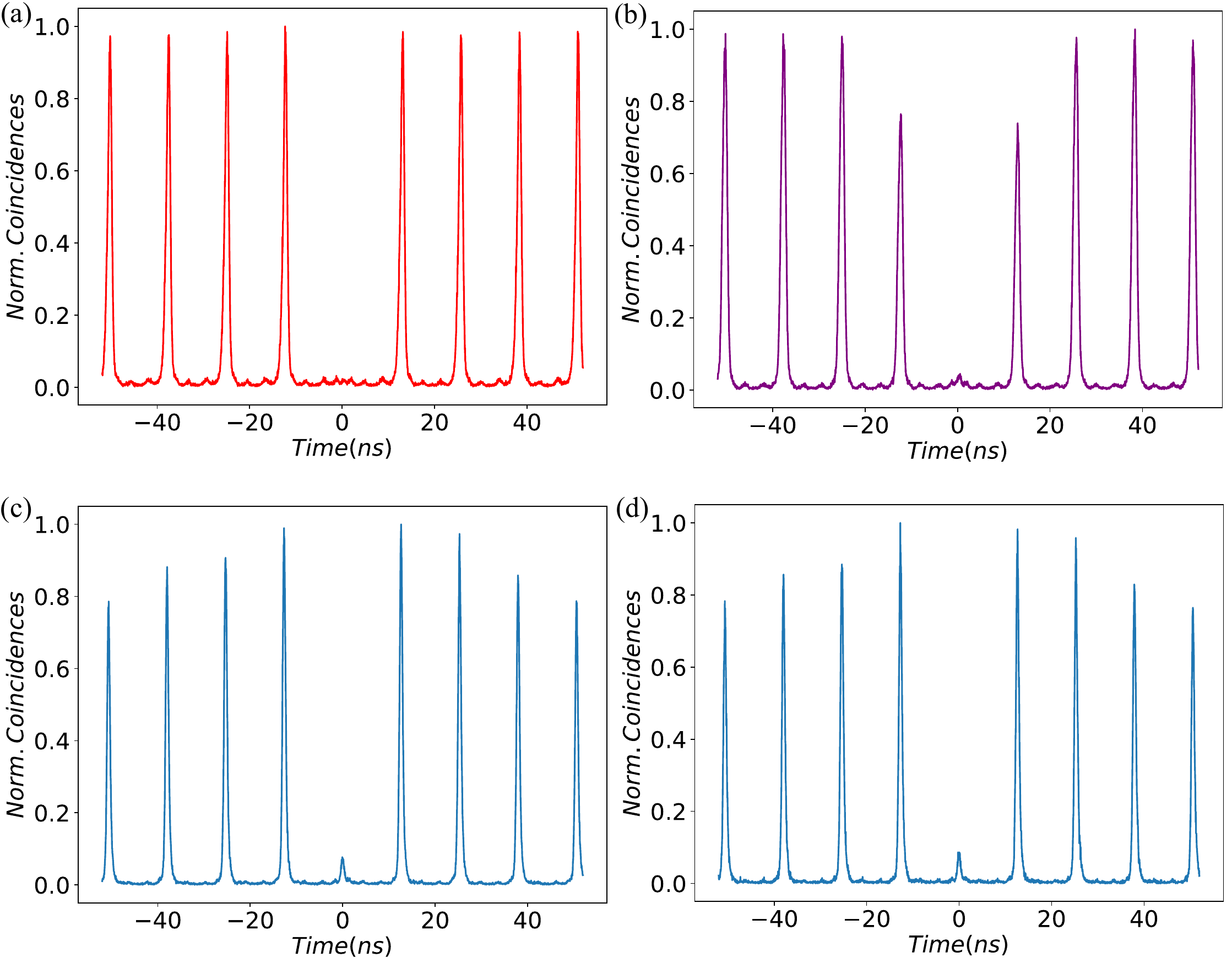}
    \caption{\textbf{HOM visibility and second-order correlation function.} (a) Measured second-order autocorrelation
histogram of our QD-based source as a function of the delay.
We obtain a single-photon purity of $g^{(2)}=(1.7 \pm 0.1)\%$. (b) Measured degree of indistinguishability for the QD-based source as a function of the delay. The obtained visibility is $V_{HOM}=0.926 \pm 0.002$. (c-d) HOM visibilities measured for the DMX output single photons. The histograms in panels (c) and (d) are associated with the cases in which the photons are delayed by $180$ ns and $360$ ns, respectively. The corresponding visibilities are $V_1=0.842 \pm 0.003$ and $V_2=0.815\pm 0.004$. }
    \label{fig:source}
\end{figure*}

In the present work, one Quantum Dot sample in a \textit{Quandela e-Delight} chip is employed as a single photon source operating in the acoustic phonon-assisted configuration. 
To characterize the performance of the quantum source, we measured both the purity and indistinguishability of single photons through typical Hanbury-Brown-Twiss and Hong-Ou-Mandel setups, measuring a $g^{(2)}=(1.7 \pm 0.1)\%$ and $V_{HOM}=0.926 \pm 0.002$ as reported in Fig. \ref{fig:source} a-b; we note that these were the typical values measured throughout the experiment.
To obtain the multi-photon resources required by the experiment, we adopted a commercial time-to-spatial demultiplexing setup - \textit{Quandela DMX6} to divide the input signal in $\sim 180$ ns bunches. Each bunch is coupled in three distinct single-mode fibers whose lengths are set to synchronize the different outputs temporally. We evaluated the pairwise indistinguishability between output photons, obtaining $V_1=0.842 \pm 0.003$ between photons delayed by $\sim 180$ ns and $V_2=0.815\pm 0.004$ for photons at $\sim 360$ ns of delay (Fig. \ref{fig:source} c-d).
 
\added{
\section{Critical points for the success probabilities of basic operations}
The sum and product operations share a probabilistic nature, and in our scheme are implemented via a post-selection procedure. Here we provide an analysis of the behaviour of the success probability functions around their critical points, i.e. points where the success probability of a given operation becomes zero.}

\added{\textbf{Product operation} For the product operation, the points where the success probability goes to zero are $(0, \infty)$ and $(\infty, 0)$, for which indeed the result of the operation is in an indefinite form.
Furthermore, we can provide an upper limit for the probability of success around the critical point. Note that, equivalently, we can define $x = \abs{z_1}$ and $y = 1/\abs{z_2}$, so that the critical point $(0, \infty)$ is mapped into $(0,0)$. With this change of variables, the success probability of the product operation can be written as
\begin{equation}
    P(x,y)  = \frac{x^2+y^2}{2(1+x^2)(1+y^2)}.
\end{equation}
Choosing a disc around the point $(0,0)$ with an arbitrarily small radius $r$, then the success probability is bounded from above as
\begin{equation}
    0 \leq P(x,y) \leq \frac{x^2+y^2}{2} \leq \frac{r^2}{2}.
\end{equation}
Hence, in the limit of multiplying a very small number times a very large number the success probability tends to $0$ by continuity:
\begin{equation}
    \lim_{(x,y) \rightarrow (0,0)} P(x,y) = P(0,0) = 0.
\end{equation}
An equivalent discussion holds for the anti-product operation.
}

\added{\textbf{Sum operation} Similarly, for both the sum and harmonic mean operations the relevant critical points are given by $(\infty, \infty)$ and $(0,0)$ respectively, corresponding to choices of inputs where the results of the operations are in an indefinite form.
If we now define $x = 1/z_1$ and $y = 1/z_2$ for the sum or $x=z_1$ and $y = z_2$ for the harmonic mean, then in both scenarios the success probability can be written as
\begin{equation}
    P(x,y) = \frac{\abs{x+y}^2+\abs{2xy}^2}{16(1+\abs{x}^2)(1+\abs{y}^2)} = \frac{\abs{x}^2+\abs{y}^2+2\abs{xy}cos (\phi_x-\phi_y) + 4 \abs{x}^2\abs{y}^2}{16 (1+\abs{x^2})(1+\abs{y^2})}.
\end{equation}
where the critical point is in $(0,0)$, in both cases.
Again, by choosing a disc around the point $(0,0)$ with an arbitrarily small radius $r$, then the success probability is bounded from above as
\begin{equation}
    0 \leq P(x,y) \leq \frac{\abs{x}^2+\abs{y}^2+2\abs{x}\abs{y}+4\abs{x}^2\abs{y}^2}{16} \leq \frac{3r^2+4r^4}{16} = \frac{r^2}{16}(3+4r^2)
    \underset{(x,y) \rightarrow (0,0) }{\longrightarrow} 0.
\end{equation}
}

\added{
\section{Upper Bound on the required number of operations}
In this section we provide a formal demonstration about the upper bound on the number of required operations - i.e. primitive field operations - for the implementation of any simulable function in the Quantum-to-Quantum Bernoulli factory paradigm. Since the space of simulable functions is given by the set of rational complex functions, defined as:
\begin{equation}
    f(z) = \frac{\sum_{i=0}^{h} \alpha_i z^i}{\sum_{j=0}^{k} \beta_j z^j} \qquad n = \max\{h,k\}
\end{equation}
we can show that if the function has degree $n$, then the maximum number of primitive field operations required is at most $4n$.
Since formally $f(z)$ can be written as a ratio between two co-prime polynomials $P(z)$ and $Q(z)$, where at least one of them has degree $n$, we can factorize the two polynomials to obtain an expression in the form:
\begin{equation}
    f(z) = \frac{P(z)}{Q(z)} = c_0 \frac{(z-p_1) \dots (z-p_h)}{(z-q_1) \dots (z-q_k)}
\end{equation}
To implement such expression in our modular scheme, we then need at most $2n$ qubits in the state $\ket{z}$ and $2n + 1$ qubits describing the root states $\ket{-p_1}, \dots, \ket{-p_n}, \ket{-q_1}, \dots, \ket{-q_n}$ and the external coefficient $\ket{c_0}$. Then, considering that the inversion operation $\ket{z} \rightarrow \ket{1/z}$ is computationally free - in the sense of not requiring extra auxiliary photons or further post-selection, to perform the implementation of each of the two polynomials we would need $n$ sums and $n-1$ products, with a pair of further product operations to combine them and multiply the result for the external coefficient $c_0$, for a total of at most $4n$ operations.
}

\section{Modeling the Partial distinguishability of the photons}
The degree of distinguishability between the input photons is mainly due to fluctuation at the quantum source level and is related to the difference in other hidden degrees of freedom between photons generated at different times, e.g. spectral wandering of the single photon emission. This translates into the generation of slightly different quantum states, reducing the fidelity between the generated state and the target one. Therefore, it is necessary to provide a theoretical model to derive an expression of the density matrix as a function of the overlap $V$ between the wave functions of the input photon pairs. In this section, we will report the corrected density matrices in the output of both building blocks, sum and product interferometer but also for the sum-product concatenation in which three photons are involved.

In what follows we will show how to take into account the partial distinguishability amongst degree of freedom which are different from the polarization one, which is exploited to encode the resource states in the QQBF, estimated by a direct measurement of the Hong-Ou-Mandel (HOM) effect. Let us assume that the single photon state emitted by the quantum dot source can be written as:
\begin{equation}
    \rho = \dyad{p} \otimes \rho_n
\end{equation}
where $\dyad{p}$ indicates a pure state in the polarization degree of freedom, and $\rho_n$ is the density matrix describing random fluctuations in all other d.o.f., e.g. spectral wandering or finite lifetime of the emission. Let us assume that $\rho_n$ can be written in its diagonal basis as:
\begin{equation}
    \rho_n = \sum_i q_i \dyad{i}
\end{equation}
with $q_0 \gg q_1 \approx q_2 \approx \ldots \approx q_n$, an assumption justified by the fact that empirically a Quantum Dot generates its photons mainly in a principal component plus additional orthogonal fluctuations. Note that the coefficients $p_i$ are related to the measured HOM visibility amongst two states $\rho_1$ and $\rho_2$ that have same polarization $\dyad{p_1} \equiv \dyad{p_2}$ as:
\begin{equation}
    V = \Tr{\rho_1 \rho_2} = \Tr{\dyad{p}^{\otimes 2}}\Tr{\rho_n^{\otimes 2}} = \sum_i q_i^2 \approx q_0 ^2
\end{equation}
in a scenario in which the two interfering photons are uncorrelated.
    
In the two photon scenario, we can then model the initial state as:
\begin{equation}
    \rho = \rho_1 \otimes \rho_2 = \rho_{\text{pol}} \otimes (p_{XX} \rho_{XX} + p_{XY} \rho_{XY})
\end{equation}
where $\rho_{XX}$ and $\rho_{XY}$ can be written as:
\begin{eqnarray}
    \rho_{XX} &=& \frac{1}{p_{XX}} \sum_i q_i^2 \dyad{i} \otimes \dyad{i} \\
    \rho_{XY} &=& \frac{1}{p_{XY}} \sum_{i \neq j} q_i q_j \dyad{i} \otimes \dyad{j} = \frac{1}{p_{XY}} \left( \sum_{i,j} q_i q_j \dyad{i} \otimes \dyad{j} - \sum_{i} q_i^2 \dyad{i} \otimes \dyad{i} \right )  
\end{eqnarray}
then the (normalization) coefficients $p_{XX}$ and $p_{XY}$ can be written as:
\begin{eqnarray}
    p_{XX} &=& \sum_i q_i^2 = V \\
    p_{XY} &=& \sum_i q_i \sum_j q_j - \sum_i q_i^2 = 1 - V
\end{eqnarray}
In such a way, the experiment can be simulated by letting the initial state $\rho$ evolve within the interferometer, post-select on the scenario where the two photons exit the interferometer separately in the trigger and output ports, and finally trace out the other degree of freedom $\rho_n$.

By using this modelization, the output density functions as a function of $V$ for the \textbf{product interferometer} are:
    \begin{itemize}
        \item Regular product :
         \begin{equation}\rho_o = N \begin{pmatrix} 
                    \abs{z_1z_2}^2 & z_1z_2V \\
                    z_1^*z_2^*V & 1\\
                \end{pmatrix}
        \end{equation}
        \item Anti-product :
        \begin{equation}\rho_o = N \begin{pmatrix} 
                    \abs{z_1z_2}^2 & -z_1z_2V \\
                    -z_1^*z_2^*V & 1\\
                \end{pmatrix}
        \end{equation}
    \end{itemize}
where $N$ is the normalization factor. In this case, it is possible to observe that the partial distinguishability affects only the off-diagonal elements introducing decoherence effects.

For the \textbf{sum interferometer} the output density function are:
    \begin{itemize}
        \item Arithmetic mean:
                \begin{equation}\rho_o = N\begin{pmatrix} 
                                \abs{z_1+z_2}^2-(z_1z_2^*+z_1^*z_2)(1-V) & (1+V)(z_1+z_2) \\
                                (1+V)(z_1^*+z_2^*) & 2(1+V)\\
                            \end{pmatrix}
                \end{equation}
        \item Harmonic mean:
                \begin{equation}\rho_o = N\begin{pmatrix} 
                                 2(1+V)\abs{z_1z_2}^2 & (1+V)z_1z_2(z_1^*+z_2^*) \\
                                 (1+V)z_1^*z_2^*(z_1+z_2) & \abs{z_1+z_2}^2-(z_1z_2^*+z_1^*z_2)(1-V)\\
                            \end{pmatrix}
                \end{equation}
    \end{itemize}
where $N$ is the normalization factor. A dependence of all matrix elements on partial distinguishability can be observed. Therefore, the addition operation shows a stronger dependence on the partial distinguishability than the product operation.

Let us consider the evaluation of the sum-product concatenation in which a three-photon experiment is performed. In this case, it is necessary in principle to consider the partial distinguishability between each pair of photons at the output of the DMX. The measured overlap decreases by increasing the temporal distance between photons; thus, output pairs are characterized by different $V_{ij}$. In particular, for the photons delayed by $\sim 180$ ns, we measured $V_1=0.842 \pm 0.003$, while the photons delayed by $\sim 360$ ns are characterized by a degree of indistinguishability of $V_2=0.815\pm 0.004$. For the sake of simplicity, in our theoretical model we assume that each pair of photons has the same degree of indistinguishability $V=0.84$. Extending the previous reasoning to this three-photon scenario, we have that the initial state can be written as:
\begin{equation}
    \rho = \rho_1 \otimes \rho_2 \otimes \rho_3 = \rho_{\text{pol}} \otimes (p_{1} \rho_{XXX} + 3 p_{2} (\rho_{XXY} + \rho_{XYX} + \rho_{YXX}) + p_{3} \rho_{XYZ})
\end{equation}
where similarly the coefficients $p_1$ and $p_2$ can be extracted from the normalization of the associated states:
\begin{eqnarray}
    \rho_{XXX} &=& \frac{1}{p_{1}} \sum_i q_i^3 \dyad{i} \otimes \dyad{i} \otimes \dyad{i}\\
    \rho_{XXY} &=& \frac{1}{p_{2}} \sum_{i \neq j} q_i^2 q_j \dyad{i} \otimes \dyad{i} \otimes \dyad{j} = \frac{1}{p_{2}} \left( \sum_{i,j} q_i^2 q_j \dyad{i} \otimes \dyad{i} \otimes \dyad{j} - \sum_{i} q_i^3 \dyad{i} \otimes \dyad{i} \otimes \dyad{i} \right )  
\end{eqnarray}
which gives:
\begin{eqnarray}
    p_{1} &=& \sum_i q_i^3 \approx q_0^3 = V^{3/2} \\
    p_{2} &=& \sum_i q_i^2 - \sum_i q_i^3 = V - V^{3/2} \\
    p_{3} &=& 1 - p_1 - 3p_2 = 1 - V^{3/2} - 3V - 3V^{3/2}
\end{eqnarray}
By computing how the states evolve within the chained interferometers and tracing out the other degrees of freedom, we obtain:
\begin{itemize}
        \item Sum Product:
                \begin{equation}\rho_o = N\begin{pmatrix} 
                                \abs{z_3}^2((z_1+V z_2)z_1^*+( V z_1+z_2)z_2^*) & (1+\sqrt{V})V(z_1+z_2)z_3 \\
                                (1+\sqrt{V})V(z_1^*+z_2^*)z_3^* & 2(1+V)\\
                            \end{pmatrix}
                \end{equation}
        \item Harmonic mean Product:
                \begin{equation}\rho_o = N\begin{pmatrix} 
                                 2(1+V)\abs{z_1z_2z_3}^2 & (1+\sqrt{V}) V (z_1^*+z_2^*)(z_3z_1z_2)\\
                                 (1+\sqrt{V})V(z_1+z_2)(z_1z_2z_3)^* & ((z_1+Vz_2)z_1^*+(V z_1+z_2)z_2^*)\\
                            \end{pmatrix}
                \end{equation}
        \item Sum Anti-product:
                \begin{equation}\rho_o = N\begin{pmatrix} 
                                \abs{z_3}^2((z_1+V z_2)z_1^*+( V z_1+z_2)z_2^*) & -(1+\sqrt{V})V(z_1+z_2)z_3 \\
                       -(1+\sqrt{V})V(z_1^*+z_2^*)z_3^* & 2(1+V)\\
                            \end{pmatrix}
                \end{equation}
        \item Harmonic mean Anti-product:
                \begin{equation}\rho_o = N\begin{pmatrix} 
                                 2(1+V)\abs{z_1z_2z_3}^2 & -(1+\sqrt{V}) Vz_1z_2(z_1^*+z_2^*)z_3\\ 
                                 -(1+\sqrt{V})V(z_1+z_2)(z_1z_2z_3)^* & ((z_1+Vz_2)z_1^*+(V z_1+z_2)z_2^*)\\
                            \end{pmatrix}
                \end{equation}
    \end{itemize}
where here $N$ indicates indicates a normalization constant.
\begin{figure}[ht]
    \centering
    \includegraphics[width=1\linewidth]{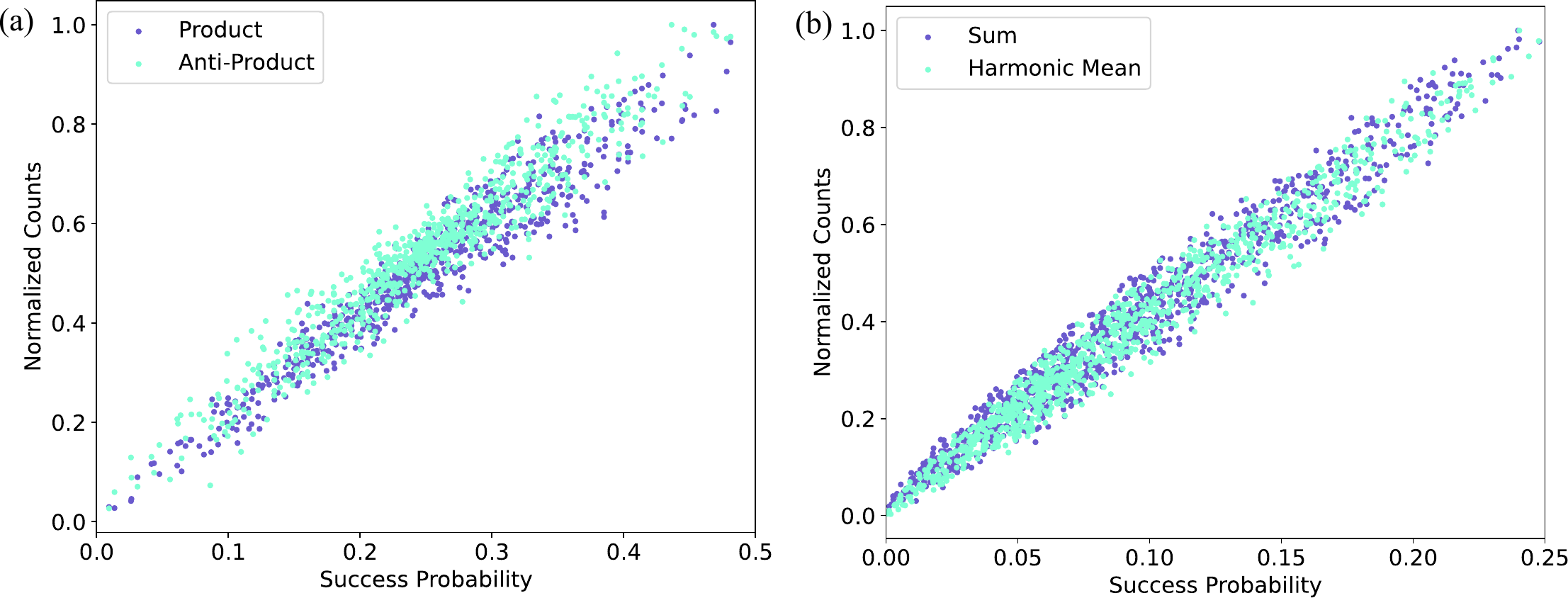}
    \caption{\textbf{Success Probability.} (a) Normalized counts collected for the product and anti-product operations as a function of the success probabilities $P_{\pm}$. All the probabilities are contained in the interval $[0,0.5]$. (b) Normalized counts collected for the arithmetic and harmonic mean operations as a function of the success probabilities $P_{S/I}$. All the probabilities are contained in the interval $[0,0.25]$.}
    \label{fig:prob_counts}
\end{figure}

\added{
\section{Uniformly sampling from the Bloch Sphere}
A general quantum state of a qubit can be parametrized as follows:
\begin{equation}
    \ket{\theta, \phi} =  \cos\frac{\theta}{2} \ket{0} + \sin \frac{\theta}{2} e^{i\phi} \ket{1}
    \label{eq:bloch}
\end{equation}
where $\theta \in [0,\pi]$ and $\phi \in [0,2\pi)$ correspond to spherical coordinates over the so-called Bloch sphere.
To validate our experimental platform, we draw randomly states $\ket{z_1}$ and $\ket{z_2}$ living in this space. In order to do so, we uniformly sample the distribution of all the possible qubit states 
employing a group isomorphism between the 2-dimensional unitary transformations space and the rotation of the unitary sphere: this procedure ensures a direct equivalence between the Haar measure of the Hilbert space and the Haar measure of the Bloch's sphere, thus establishing a link between the uniform distributions over the two spaces.
In the Bloch sphere picture, since the parameters $(\theta, \phi)$ are associated with the polar angles in spherical coordinates, to extract a random quantum state one can draw from the space of points lying on the surface of a unitary-radius sphere. Therefore, in order to sample this space uniformly, one can equivalently sample the polar parameters as:
\begin{equation}
    \begin{split}
        \phi \in U(0,2\pi) \\
        \theta \in \arccos (U(-1,1))
    \end{split}
\end{equation}
that is in the rectangle generated by the variable $\phi \in [0,2\pi)$ and $h \in (-1,1]$, where $h = \cos \theta$ and $U(\alpha,\beta)$ indicates a uniformly distributed real-valued variable between $\alpha$ and $\beta$.
As in the main text, our input states are written as:
\begin{equation}
    \ket{z} = \frac{z\ket{0}+\ket{1}}{\sqrt{1+\abs{z}^2}}
\end{equation}
Thus, according to Eq. \eqref{eq:bloch}, the complex parameter $z$ can be directly related to the parameters associated with the Bloch sphere as:
\begin{equation}
    z = \cot \frac{\theta}{2}e^{-i\phi} = \sqrt{\frac{1+\cos \theta}{1-\cos \theta}} e^{-i\phi}
\end{equation}
Then, in order to generate the complex parameter $z$ in a way that the associated states follow a uniform distribution over the two-dimensional Hilbert space we can use the following formula
\begin{equation}
    z = \sqrt{\frac{1+h}{1-h}} e^{-i\phi} \qquad (h,\phi) \in (-1,1]\cross[0,2\pi) \; \; \text{uniformly}. 
\end{equation}
}

\section{Experimental Details - Three photon operation}

As reported in Section 'Polarization-based implementation of a photonic Quantum-to-Quantum Bernoulli factory' of the main text, the field operators, that is,sum and product, are implemented via probabilistic interferometric schemes. The success probabilities associated with the product interferometer have the same expression for both product and anti-product operations, and the maximum value reachable is $P_{\pm}^{max}=0.5$. In Fig. \ref{fig:prob_counts} a, we report the normalized counts collected for each experimental point with respect to the $P_{\pm} (z_1,z_2)$ computed following the theoretical expression. As expected, a higher success probability corresponds to higher experimental counts, with an observed linear dependence.  

Although the success probabilities associated with the arithmetic and harmonic mean have different expressions, both of them can be at most equal to $P_{S/I}^{max}=0.25$. Also in this case, by reporting the collected normalized counts as a function of the computed success probability we observe a linear trend for both operations (Fig. \ref{fig:prob_counts} b).

Let us consider the concatenation of the sum and product interferometers. In this case, four different combined operations can be evaluated, i.e. sum-product, harmonic mean-product, sum-anti-product, and harmonic mean-anti-product. The corresponding success probabilities can be directly computed as the product of the combined $P_{\pm}$ and $P_{S/I}$.
\begin{eqnarray}
    P_{SP}&=&P_{+}((z_1+z_2)/2,z_3)P_{S}(z_1,z_2)\\
    P_{MP}&=&P_{+}(2z_1 z_2/(z_1+z_2),z_3)P_{I}(z_1,z_2)\\
    P_{SA}&=&P_{-}((z_1+z_2)/2,z_3)P_{S}(z_1,z_2)\\
    P_{MA}&=&P_{-}(2z_1 z_2/(z_1+z_2),z_3)P_{I}(z_1,z_2)
\end{eqnarray}
To demonstrate the modularity of the proposed approach, we experimentally evaluated the outputs of the sum-product concatenation for $30$ different sets of inputs $(z_1,z_2,z_3)$. In Table \ref{Tab:fid_3f} we report the four success probabilities for each input set and the fidelity computed assuming both for perfect and partial distinguishability.

\begin{figure*}[hb!]
    \centering
    \includegraphics[width=1\linewidth]{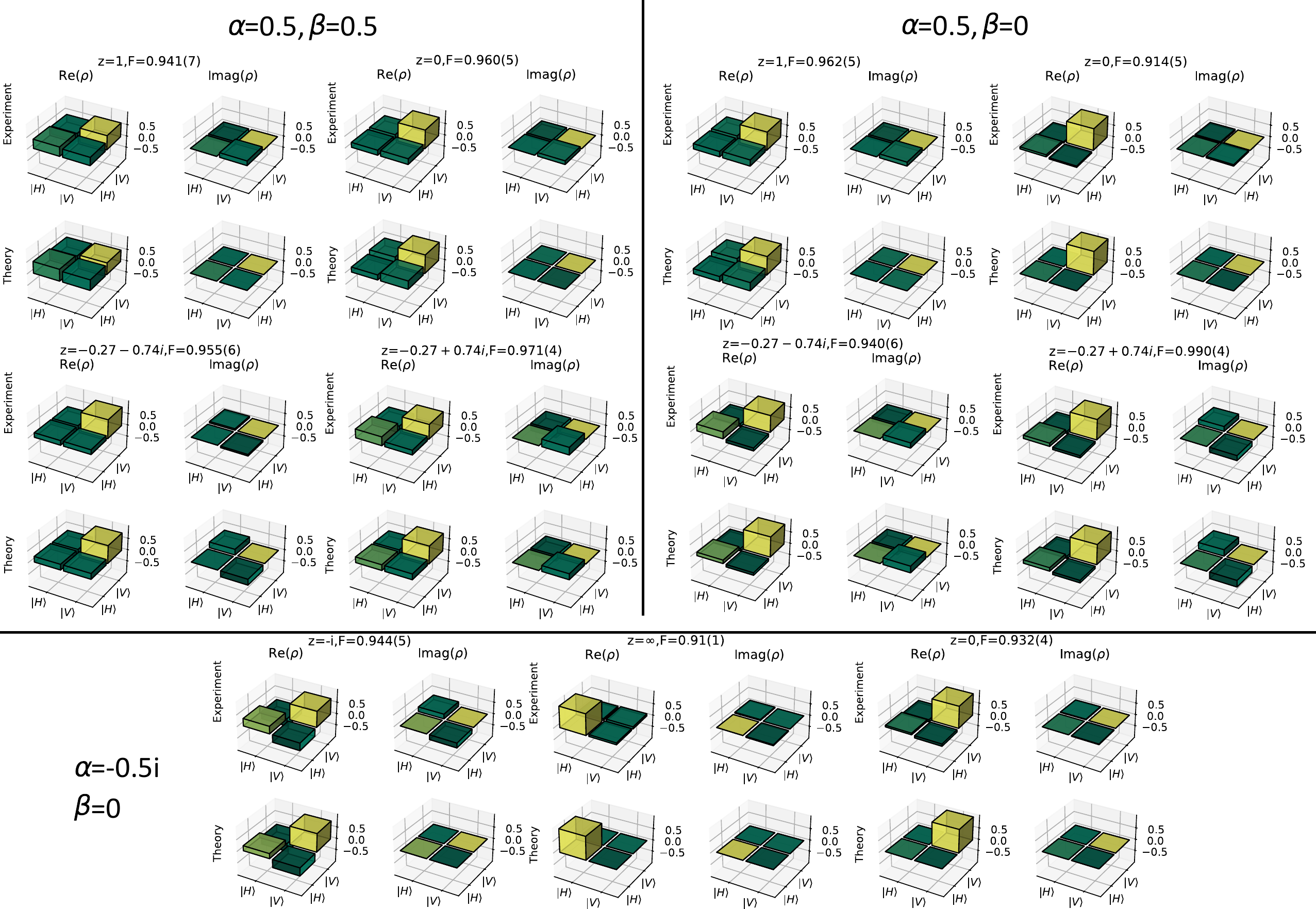}
    \caption{\textbf{Experimentally obtained output density matrices of the states reported in Fig. 4(b) of the Main Text.} We report the density matrices reconstructed experimentally via Quantum State Tomography, related to the states presented in Fig. 4 (b) of the Main Text. In particular, we compute the output function $\ket{g_{\alpha \beta}(z)} = \ket{\alpha z + \beta}$ on several data states $\ket{z}$ with the parameters experimentally tuned to $(\alpha, \beta)=\{(1,1), (0,1), (0,-1)\}$. In the figure, we depict a comparison between the experimentally produced output states on top and the theoretically predicted states, corrected for the measured limited HOM visibility $V \sim 84 \%$ on the bottom.}
    \label{fig:tomos}
\end{figure*}

In Figure \ref{fig:tomos}, we report the comparison between the expected density matrices at the output of the concatenated sum and product modules and the experimentally obtained ones; for the states reported in Fig. 4b of the Main Text.
\newpage
\begin{longtable*}{c|c|c|c|c}
\caption{\textbf{Sum-product concatenation.} For each of the $30$ set of input states $(z_1,z_2,z_3)$ we report (i) in the first row, the success probabilities for the four combined operations, (ii) in the second row, the fidelities computed under the hypothesis of perfect indistinguishability, and (iii) in the third row, the fidelities obtained assuming partial distinguishability $V=0.84$.}\\ \label{Tab:fid_3f} 
\multirow{3}{*}{ (1,1,1) } & $0.031$ & $0.031$ & $0.031$ & $0.031$ \\ \cline{2-5}  & $0.846\pm0.011$ & $0.843\pm0.015$ & $0.867\pm0.010$ & $0.854\pm0.012$ \\ \cline{2-5}  & $0.941\pm0.007$ & $0.935\pm0.006$ & $0.893\pm0.011$ & $0.894\pm0.014$ \\ \hline
\multirow{3}{*}{ (1,-1,-1) } & $0.016$ & $0.016$ & $0.016$ & $0.016$ \\ \cline{2-5}  & $0.769\pm0.014$ & $0.928\pm0.009$ & $0.773\pm0.015$ & $0.848\pm0.015$ \\ \cline{2-5}  & $0.949\pm0.010$ & $0.996\pm0.003$ & $0.941\pm0.006$ & $0.972\pm0.005$ \\ \hline
\multirow{3}{*}{ (-1,-1,1) } & $0.031$ & $0.031$ & $0.031$ & $0.031$ \\ \cline{2-5}  & $0.846\pm0.009$ & $0.870\pm0.008$ & $0.906\pm0.007$ & $0.899\pm0.009$ \\ \cline{2-5}  & $0.952\pm0.006$ & $0.957\pm0.005$ & $0.955\pm0.007$ & $0.952\pm0.006$ \\ \hline
\multirow{3}{*}{ (-i,i,-i) } & $0.016$ & $0.016$ & $0.016$ & $0.016$ \\ \cline{2-5}  & $0.821\pm0.015$ & $0.885\pm0.013$ & $0.758\pm0.017$ & $0.915\pm0.010$ \\ \cline{2-5}  & $0.975\pm0.004$ & $0.984\pm0.004$ & $0.938\pm0.008$ & $0.996\pm0.002$ \\ \hline
\multirow{3}{*}{ (-i,-i,i) } & $0.031$ & $0.031$ & $0.031$ & $0.031$ \\ \cline{2-5}  & $0.858\pm0.009$ & $0.864\pm0.010$ & $0.811\pm0.011$ & $0.795\pm0.013$ \\ \cline{2-5}  & $0.912\pm0.007$ & $0.916\pm0.010$ & $0.899\pm0.008$ & $0.900\pm0.009$ \\ \hline
\multirow{3}{*}{ (-i,0,-i) } & $0.039$ & $0.008$ & $0.039$ & $0.008$ \\ \cline{2-5}  & $0.867\pm0.011$ & $0.817\pm0.019$ & $0.836\pm0.010$ & $0.824\pm0.023$ \\ \cline{2-5}  & $0.944\pm0.005$ & $0.821\pm0.022$ & $0.916\pm0.006$ & $0.811\pm0.031$ \\ \hline
\multirow{3}{*}{ (0,i,-i) } & $0.039$ & $0.008$ & $0.039$ & $0.008$ \\ \cline{2-5}  & $0.859\pm0.012$ & $0.886\pm0.013$ & $0.911\pm0.013$ & $0.863\pm0.015$ \\ \cline{2-5}  & $0.967\pm0.004$ & $0.889\pm0.012$ & $0.988\pm0.003$ & $0.871\pm0.016$ \\ \hline
\multirow{3}{*}{ ($\infty$,0,-i) } & $0.016$ & $0.016$ & $0.016$ & $0.016$ \\ \cline{2-5}  & $0.908\pm0.011$ & $0.788\pm0.016$ & $0.923\pm0.010$ & $0.797\pm0.020$ \\ \cline{2-5}  & $0.910\pm0.010$ & $0.781\pm0.015$ & $0.919\pm0.010$ & $0.802\pm0.016$ \\ \hline
\multirow{3}{*}{ (1,0,1) } & $0.039$ & $0.008$ & $0.039$ & $0.008$ \\ \cline{2-5}  & $0.873\pm0.010$ & $0.869\pm0.023$ & $0.870\pm0.009$ & $0.818\pm0.025$ \\ \cline{2-5}  & $0.962\pm0.005$ & $0.865\pm0.021$ & $0.953\pm0.004$ & $0.822\pm0.021$ \\ \hline
\multirow{3}{*}{ (0,1,1) } & $0.039$ & $0.008$ & $0.039$ & $0.008$ \\ \cline{2-5}  & $0.866\pm0.009$ & $0.881\pm0.014$ & $0.894\pm0.013$ & $0.880\pm0.018$ \\ \cline{2-5}  & $0.960\pm0.005$ & $0.885\pm0.018$ & $0.969\pm0.004$ & $0.874\pm0.016$ \\ \hline
\multirow{3}{*}{ ($\infty$,0,1) } & $0.016$ & $0.016$ & $0.016$ & $0.016$ \\ \cline{2-5}  & $0.911\pm0.008$ & $0.807\pm0.016$ & $0.913\pm0.013$ & $0.839\pm0.015$ \\ \cline{2-5}  & $0.908\pm0.010$ & $0.811\pm0.014$ & $0.910\pm0.014$ & $0.840\pm0.011$ \\ \hline
\multirow{3}{*}{ (0,0,-i) } & $0.063$ & $0.000$ & $0.063$ & $0.000$ \\ \cline{2-5}  & $0.932\pm0.004$ & $0.000\pm0$ & $0.906\pm0.006$ & $0.000\pm0$ \\ \cline{2-5}  & $0.932\pm0.004$ & $0.000\pm0$ & $0.909\pm0.008$ & $0.000\pm0$ \\ \hline
\multirow{3}{*}{ (0,0,1) } & $0.062$ & $0.000$ & $0.062$ & $0.000$ \\ \cline{2-5}  & $0.915\pm0.006$ & $0.000\pm0$ & $0.923\pm0.005$ & $0.000\pm0$ \\ \cline{2-5}  & $0.914\pm0.005$ & $0.000\pm0$ & $0.922\pm0.004$ & $0.000\pm0$ \\ \hline
\multirow{3}{*}{ (0,0,0) } & $0.125$ & $0.000$ & $0.125$ & $0.000$ \\ \cline{2-5}  & $0.987\pm0.005$ & $0.000\pm0$ & $0.989\pm0.003$ & $0.000\pm0$ \\ \cline{2-5}  & $0.986\pm0.005$ & $0.000\pm0$ & $0.989\pm0.003$ & $0.000\pm0$ \\ \hline
\multirow{3}{*}{ ($\infty$,$\infty$,$\infty$) } & $0.000$ & $0.125$ & $0.000$ & $0.125$  \\ \cline{2-5}  & $0.000\pm0$ & $0.980\pm0.002$ & $0.000\pm0$ & $0.983\pm0.002$ \\ \cline{2-5}  & $0.000\pm0$ & $0.980\pm0.002$ & $0.000\pm0$ & $0.983\pm0.002$ \\ \hline
\multirow{3}{*}{ (-0.13+0.88i,-0.24-0.15i,1) } & $0.038$ & $0.008$ & $0.038$ & $0.008$ \\ \cline{2-5}  & $0.888\pm0.008$ & $0.817\pm0.015$ & $0.875\pm0.011$ & $0.762\pm0.022$ \\ \cline{2-5}  & $0.981\pm0.004$ & $0.946\pm0.012$ & $0.975\pm0.004$ & $0.919\pm0.017$ \\ \hline
\multirow{3}{*}{ (-0.27+0.74i,0,1) } & $0.044$ & $0.006$ & $0.044$ & $0.006$ \\ \cline{2-5}  & $0.921\pm0.009$ & $0.877\pm0.017$ & $0.870\pm0.012$ & $0.858\pm0.028$ \\ \cline{2-5}  & $0.990\pm0.004$ & $0.878\pm0.025$ & $0.969\pm0.005$ & $0.853\pm0.017$ \\ \hline
\multirow{3}{*}{ (-0.27+0.74i,1,1) } & $0.024$ & $0.017$ & $0.024$ & $0.017$ \\ \cline{2-5}  & $0.818\pm0.013$ & $0.734\pm0.015$ & $0.779\pm0.012$ & $0.773\pm0.014$ \\ \cline{2-5}  & $0.955\pm0.006$ & $0.921\pm0.009$ & $0.935\pm0.008$ & $0.951\pm0.007$ \\ \hline
\multirow{3}{*}{ (0,-0.27-0.74i,1) } & $0.044$ & $0.006$ & $0.044$ & $0.006$ \\ \cline{2-5}  & $0.822\pm0.011$ & $0.825\pm0.016$ & $0.833\pm0.012$ & $0.822\pm0.024$ \\ \cline{2-5}  & $0.940\pm0.006$ & $0.824\pm0.022$ & $0.946\pm0.007$ & $0.827\pm0.020$ \\ \hline
\multirow{3}{*}{ (1,-0.27-0.74i,1) } & $0.024$ & $0.017$ & $0.024$ & $0.017$ \\ \cline{2-5}  & $0.872\pm0.014$ & $0.915\pm0.014$ & $0.889\pm0.013$ & $0.878\pm0.013$ \\ \cline{2-5}  & $0.971\pm0.004$ & $0.986\pm0.005$ & $0.971\pm0.005$ & $0.987\pm0.004$ \\ \hline
\multirow{3}{*}{ (-0.27+0.74i,0,-1) } & $0.044$ & $0.006$ & $0.044$ & $0.006$ \\ \cline{2-5}  & $0.832\pm0.009$ & $0.818\pm0.030$ & $0.896\pm0.010$ & $0.861\pm0.022$ \\ \cline{2-5}  & $0.940\pm0.008$ & $0.825\pm0.024$ & $0.981\pm0.003$ & $0.867\pm0.022$ \\ \hline
\multirow{3}{*}{ (-0.27+0.74i,1,-1) } & $0.024$ & $0.017$ & $0.024$ & $0.017$ \\ \cline{2-5}  & $0.762\pm0.012$ & $0.734\pm0.019$ & $0.779\pm0.010$ & $0.730\pm0.018$ \\ \cline{2-5}  & $0.914\pm0.010$ & $0.924\pm0.010$ & $0.916\pm0.008$ & $0.909\pm0.009$ \\ \hline
\multirow{3}{*}{ (-0.31-0.40i,-3.25+0.90i,-0.53+1.15i) } & $0.019$ & $0.024$ & $0.019$ & $0.024$ \\ \cline{2-5}  & $0.927\pm0.016$ & $0.934\pm0.012$ & $0.947\pm0.019$ & $0.853\pm0.016$ \\ \cline{2-5}  & $0.958\pm0.011$ & $0.925\pm0.019$ & $0.994\pm0.003$ & $0.946\pm0.011$ \\ \hline
\multirow{3}{*}{ (1.97+2.41i,0.04+0.65i,2.03-0.71i) } & $0.024$ & $0.032$ & $0.024$ & $0.032$ \\ \cline{2-5}  & $0.908\pm0.009$ & $0.855\pm0.010$ & $0.908\pm0.013$ & $0.872\pm0.010$ \\ \cline{2-5}  & $0.946\pm0.007$ & $0.950\pm0.006$ & $0.952\pm0.007$ & $0.959\pm0.004$ \\ \hline
\multirow{3}{*}{ (0.06+0.15i,-0.33+1.77i,0.08-1.33i) } & $0.028$ & $0.011$ & $0.028$ & $0.011$ \\ \cline{2-5}  & $0.845\pm0.008$ & $0.763\pm0.012$ & $0.819\pm0.013$ & $0.830\pm0.020$ \\ \cline{2-5}  & $0.973\pm0.004$ & $0.886\pm0.013$ & $0.937\pm0.005$ & $0.917\pm0.010$ \\ \hline
\multirow{3}{*}{ (-1.09-1.77i,-0.78+1.02i,0.58-0.33i) } & $0.009$ & $0.026$ & $0.009$ & $0.026$ \\ \cline{2-5}  & $0.796\pm0.016$ & $0.894\pm0.009$ & $0.824\pm0.014$ & $0.966\pm0.009$ \\ \cline{2-5}  & $0.949\pm0.008$ & $0.992\pm0.004$ & $0.925\pm0.011$ & $0.992\pm0.006$ \\ \hline
\multirow{3}{*}{ (-0.49-0.03i,1.10-0.28i,-0.36-0.95i) } & $0.024$ & $0.010$ & $0.024$ & $0.010$ \\ \cline{2-5}  & $0.753\pm0.010$ & $0.762\pm0.019$ & $0.754\pm0.013$ & $0.786\pm0.017$ \\ \cline{2-5}  & $0.934\pm0.005$ & $0.974\pm0.005$ & $0.932\pm0.007$ & $0.955\pm0.007$ \\ \hline
\multirow{3}{*}{ (0.23-0.32i,0.95-0.78i,-1.37+0.23i) } & $0.033$ & $0.016$ & $0.033$ & $0.016$ \\ \cline{2-5}  & $0.926\pm0.009$ & $0.904\pm0.016$ & $0.906\pm0.009$ & $0.839\pm0.015$ \\ \cline{2-5}  & $0.995\pm0.002$ & $0.994\pm0.004$ & $0.995\pm0.002$ & $0.968\pm0.006$ \\ \hline
\multirow{3}{*}{ (0.12+0.06i,-1.51-0.72i,-0.20-0.25i) } & $0.031$ & $0.017$ & $0.031$ & $0.017$ \\ \cline{2-5}  & $0.918\pm0.009$ & $0.921\pm0.010$ & $0.913\pm0.010$ & $0.952\pm0.006$ \\ \cline{2-5}  & $0.978\pm0.005$ & $0.942\pm0.007$ & $0.969\pm0.005$ & $0.975\pm0.005$ \\ \hline
\multirow{3}{*}{ (-0.86-0.43i,-0.44-0.87i,0.53-0.26i) } & $0.032$ & $0.028$ & $0.032$ & $0.028$ \\ \cline{2-5}  & $0.833\pm0.012$ & $0.816\pm0.013$ & $0.835\pm0.011$ & $0.829\pm0.010$ \\ \cline{2-5}  & $0.945\pm0.006$ & $0.938\pm0.008$ & $0.909\pm0.007$ & $0.898\pm0.008$ \\ \hline

\end{longtable*}

\section{Bernoulli factory as a programmable setup}
As shown in the main text, the proposed experimental implementation of the QQBF is obtained with a modular approach. Therefore the in-bulk interferometers implementing the sum and product building blocks can be concatenated to compute even more complex functions. For simplicity, we exclude the inversion function from the following analysis because it is computationally free, in the sense of not requiring extra auxiliary photons or further post-selection.


Let us consider the QQBF as a setup that can be programmed by a user, with the goal of computing a general rational function of degree $d$, that is:
\begin{equation}
    g_{\alpha_0,\alpha_1, \ldots \alpha_d}(z)=\sum_{i=0}^d \alpha_i z^i = c_0(z - p_1) \cdot \ldots \cdot (z - p_d)
\end{equation}

In order to do so, $d$ copies of the unknown quantum state $\ket{z}$ which is to be manipulated into a quantum coin $\ket{g_{p_1, z_2,\ldots,p_d}(z)}$ are required, together with $d + 1$ photons encoded in the register states $\{\ket{-p_1}, \ldots, \ket{-p_d}, \ket{c_0}\}$ and defining the particular polynomial being implemented, are required to be fed to a circuit made up of at most $2d$ building blocks. We note that for particular polynomials, e.g. when some roots are equal to zero, more efficient circuits can be devised.
In the main text, we limited our analysis to verify the QQBF performance in three-photon experiments by concatenating two modules. In this configuration, the programmable functions depend on the order in the concatenation and on the number of inputs identifying the unknown state. In particular, by considering first sum and then product, the list of functions $g^{SP}(z)$ we can implement in this way is the following:
\begin{enumerate}
    \item[1.] $(a,z,z)$ or $ (z,a,z) \rightarrow \frac{(a+z)z}{2}=$$\alpha_2 z^2+ z/2$
    \item[2.] $(z,z,a) \rightarrow a z=\alpha_1 z$
    \item[3.] $(a,b,z)$ or $\frac{(b,a,z)}{2} \rightarrow (a+b)z=\alpha_1 z$
    \item[4.] $(a,z,b)$ or $(z,a,b) \rightarrow \frac{(z+a)b}{2}=$$\alpha_1 z+\alpha_0$.
\end{enumerate}
It is worth noting that these results are obtained upon post-selection conditioned on the detection of one photon in state $\ket{S}_{T_S}$ at the sum trigger output and one photon
in the state $\ket{+}_{T_P}$ at the product trigger output, but it is also possible to detect the second photon in the anti-product trigger output $\ket{-}_{T_P}$. In this alternative case, the functions $g^{SAP}(z)=-g^{SP}(z)$ are computed. 
Inverting the order of the concatenation, the programmable functions are:
\begin{itemize}
    \item[5.] $(a,z,z)$ or $ (z,a,z) \rightarrow \frac{(a+1)}{2}z=\alpha_1 z$ 
    \item[6.] $(z,z,a) \rightarrow \frac{z^2+\alpha}{2}=$ $z^2/2+\alpha_0$
    \item[7.] $(a,b,z)$ or $(b,a,z) \rightarrow \frac{a b+z}{2}=\alpha_0 + z/2$
    \item[8.] $(a,z,b)$ or $(a,z,b) \rightarrow \frac{a z+ b}{2}=$ $\alpha_1 z+\alpha_0$.
\end{itemize}
While functions 1., 4., 5., and 7. can be implemented using only one module, either addition or subtraction, the other functions require both modules. In certain cases, there is more than one possible configuration to construct the same function. As an example, function $\alpha_1 z+\alpha_0$, as we have seen previously, can be generated with both orders for the block scheme. The more suitable implementation can be identified via different criteria. Focusing on the success probability, the above function is generated in the sum-product and product-sum concatenations with $P_{SP}$ and $P_{PS}$, respectively:
\begin{equation}
    P_{SP}=P_{+}(2\alpha_1, z+\alpha_0/\alpha_1)P_S(z,\alpha_0/\alpha_1)=\frac{1+\abs{\alpha_1 z +\alpha_0}^2}{8(1+\abs{z}^2)(1+\abs{\alpha_0/\alpha_1}^2)(1+\abs{2\alpha_1}^2)}
\end{equation}
\begin{equation}
    P_{PS}=P_S(2\alpha_0,z 2 \alpha_1) P_{+}(z,2 \alpha 1)=\frac{1+\abs{\alpha_1 z +\alpha_0}^2}{8(1+\abs{z}^2)(1+\abs{2 \alpha_0}^2)(1+\abs{2\alpha_1}^2)}
\end{equation}
These expressions show that it is more convenient to choose the sum-product order if $\abs{\alpha_1}^2 \leq 4$, or the reverse order in case this condition is not satisfied.
